\documentclass[10pt,aps,pra,twocolumn,superscriptaddress,floatfix,showpacs,longbibliography]{revtex4-1}

\usepackage{graphicx}
\usepackage{amsfonts}
\usepackage{amssymb}
\usepackage{amsmath}
\usepackage{txfonts}
\usepackage{lipsum}
\usepackage{color}
\usepackage{wasysym}
\usepackage{hyperref}
\usepackage{bbold}
\usepackage{etoolbox} 
\usepackage[normalem]{ulem}
\usepackage{mathtools} 
\usepackage{multirow} 
\usepackage{hhline}
\usepackage{mathrsfs} 
\usepackage{soul}

\usepackage{letltxmacro}
\LetLtxMacro{\oldsqrt}{\sqrt}
\renewcommand{\sqrt}[2][\mkern8mu]{\mkern-6mu\mathop{}\oldsqrt[#1]{#2}}

\DeclareMathOperator{\Tr}{Tr}

\newcommand{\av}[1]{\ensuremath{\left\langle #1 \right\rangle}}

\begin{document}

\title{Spin dynamics of itinerant electrons: local magnetic moment formation and Berry phase}

\author{E. A. Stepanov}
\email{evgeny.stepanov@polytechnique.edu}
\affiliation{CPHT, CNRS, Ecole Polytechnique, Institut Polytechnique de Paris, F-91128 Palaiseau, France}

\author{S. Brener}
\affiliation{I. Institute of Theoretical Physics, University of Hamburg, Jungiusstrasse 9, 20355 Hamburg, Germany}
\affiliation{The Hamburg Centre for Ultrafast Imaging, Luruper Chaussee 149, 22761 Hamburg, Germany}

\author{V. Harkov}
\affiliation{I. Institute of Theoretical Physics, University of Hamburg, Jungiusstrasse 9, 20355 Hamburg, Germany} 

\affiliation{European X-Ray Free-Electron Laser Facility, Holzkoppel 4, 22869 Schenefeld, Germany}

\author{M. I. Katsnelson}
\affiliation{Radboud University, Institute for Molecules and Materials, 6525AJ Nijmegen, The Netherlands}

\author{A. I. Lichtenstein}
\affiliation{I. Institute of Theoretical Physics, University of Hamburg, Jungiusstrasse 9, 20355 Hamburg, Germany}
\affiliation{The Hamburg Centre for Ultrafast Imaging, Luruper Chaussee 149, 22761 Hamburg, Germany}
\affiliation{European X-Ray Free-Electron Laser Facility, Holzkoppel 4, 22869 Schenefeld, Germany}

\begin{abstract}
The {\it state-of-the-art} theoretical description of magnetic materials relies on solving effective Heisenberg spin problems or their generalizations to relativistic or multi-spin-interaction cases that explicitly assume the presence of local magnetic moments in the system. 
We start with a general interacting fermionic model that is often obtained in {\it ab initio} electronic structure calculations and show that the corresponding spin problem can be introduced even in the paramagnetic regime, which is characterized by a zero average value of the magnetization.
Further, we derive a physical criterion for the formation of the local magnetic moment and confirm that the latter exists already at high temperatures well above the transition to the ordered magnetic state.
The use of path-integral techniques allows us to disentangle spin and electronic degrees of freedom and to carefully separate rotational dynamics of the local magnetic moment from Higgs fluctuations of its absolute value.
It also allows us to accurately derive the topological Berry phase and relate it to a physical bosonic variable that describes dynamics of the spin degrees of freedom.
As the result, we demonstrate that the equation of motion in the case of a large magnetic moment takes a conventional Landau-Lifshitz form that explicitly accounts for the Gilbert damping due to itinerant nature of the original electronic model.
\end{abstract}

\maketitle

\section{Introduction}

The problem of magnetism of itinerant electrons~\cite{Herring,M2,Moriya_book,IKS_book} seems to be one of the most challenging among all complicated many-body problems relevant for condensed matter. 
The key point is the duality between itinerant and localized (atomic-like) behavior of $d$-electrons in magnetic transition metals and their alloys. 
To make the long story (described in the books cited above) short, direct measurements of Fermi surfaces and transport properties of iron-group metals clearly indicate itinerant properties of $3d$-electrons whereas temperature dependence of the magnetic susceptibility (Curie-Weiss law) and neutron scattering data are reliable evidences of localized behavior of magnetic moments, both below and above the Curie temperatures. 
This itinerant-localized duality can be considered as a bright manifestation of generic particle-wave duality in quantum many-body systems \cite{VKT1993} and thus has a general physical interest. 

The finite-temperature state with local magnetic moments cannot be described within Landau Fermi-liquid theory since it violates the main Landau postulate on one-to-one correspondence between the states of bare particles and quasiparticles~\cite{VK1989}. 
The combination of itinerant band theory based on a density functional with atomic-like effective impurity approach based on dynamical mean-field theory (DMFT)~\cite{RevModPhys.68.13} allows one to describe quantitatively this duality and the properties of the iron-group metals \cite{KL1999,LiKatKot}. 
However, this approach does not solve the problem completely, not least because the non-local correlation effects beyond DMFT seem to play a crucial role in elemental iron \cite{Fe_beyond_DMFT}.

Describing the magnetic properties of itinerant systems with local magnetic moments from the point of view of a generic quantum many-body theory is quite problematic. 
For localized spins, the most convenient technique is based on the use of path integrals over spin coherent states \cite{Inomata_book,Auerbach_book}. 
Evaluating the path integrals by the saddle-point method leads to the classical equation of the spin precession, the ``kinetic'' term with the first derivative in time originating from the topological (Berry) phase \cite{Schapere_book}. 
For this construction, conservation of the length of the total spin on each site is crucially important. 
In itinerant-electron systems, the length of local magnetic moment is fluctuating which changes dramatically the mathematical structure of the theory.

In many contemporary works, the problem of fluctuating length of the local magnetic moment is described in terms of the ``Higgs mode''~\cite{doi:10.1146/annurev-conmatphys-031214-014350} borrowed from high-energy physics~\cite{PhysRevLett.13.321, HIGGS1964132, PhysRevLett.13.508, PhysRevLett.13.585}.
Indeed, whereas magnons associated to the spin rotations are Goldstone modes originating from the broken rotational invariance in spin space, the fluctuating length of the spin is naturally associated to the massive Higgs mode. 
The magnetic Higgs mode has been observed in quantum dimer systems~\cite{PhysRevLett.100.205701, merchant2014quantum}.
Recent inelastic neutron scattering experiments have also revealed Higgs oscillations of the magnetization in Ca$_2$RuO$_4$~\cite{jain2017higgs, PhysRevLett.119.067201} and C$_9$H$_{18}$N$_2$CuBr$_4$~\cite{hong2017higgs, PhysRevLett.122.127201}.

Theoretically, Higgs fluctuations that are seen in experiments are usually explained on a basis of spin~\cite{jain2017higgs, hong2017higgs} or bosonic Hubbard~\cite{endres2012higgs, PhysRevLett.109.010401, PhysRevLett.110.170403} models.
However, for fermionic systems which we are interested in here, it was not possible to introduce a proper Higgs field that couples to electrons. 
Usually, this field was introduced via decoupling of the interaction term~\cite{sachdev2008quantum, PhysRevX.8.011012, ScheurerE3665, PhysRevX.8.021048, PhysRevX.10.041057}.
The problem with this trick is that such field has no clear physical meaning whereas the free energy of the system is always plotted as a function of this field and not of the observed quantity, such as magnetization.
The second problem is that this field is frequently treated in a mean-field approximation assuming that the magnetization (and the field) has a non-zero average. This approach is problematic (or even useless) in paramagnetic phase. In other words, it is not possible to prove that this field can be connected to a local magnetic moment, because the local magnetization does not necessarily point in the direction of the field.  

Another problem is to derive the equation of motion for rotational dynamics of the {\it itinerant-electron} spin. 
It is not easy at all to generalize the formalism of spin-coherent states and related Berry phase for fermionic systems. 
Usually, one introduces rotation matrices for a quantization axis of fermions to get a Berry phase~\cite{PhysRevLett.65.2462, PhysRevB.43.3790, doi:10.1142/S0217979200002430, DUPUIS2001617}. 
However, previously it was not possible to connect this Berry phase to a proper bosonic variable that describes the modulus of magnetization.
In addition, in paramagnetic situation one cannot claim that the variation of magnetization angles are small, which is usually necessary for these calculations.
The difficulty of introducing physical variables for describing the rotational dynamics of the local magnetization is directly related to the fact that spin degrees of freedom that can be associated with the local magnetic moment cannot be easily separated from electronic degrees of freedom that form this local moment. 
Therefore, the equation of motion in the presence of electrons was possible to derive (until now) only assuming the existence of a classical spin~\cite{Sayad_2015, PhysRevLett.117.127201}.

In this paper, we solve this problem based on a modification and development~\cite{PhysRevLett.121.037204, PhysRevB.99.115124, PhysRevB.100.205115, PhysRevB.103.245123} of the dual boson formalism~\cite{Rubtsov20121320, PhysRevB.90.235135, PhysRevB.93.045107, PhysRevB.94.205110, PhysRevB.100.165128}. 
Our aim is twofold: to derive properly the Berry phase term describing spin precession for fermionic (that is, itinerant) systems in terms of the physical bosonic variables, and to write a physical criterion for the formation of local magnetic moments. Our approach is applicable for both magnetically ordered and paramagnetic phases and does not assume the smallness of the rotation angles, the standard assumption in the theory of exchange interactions in magnetically ordered phases~\cite{LKG84, LKG85, LKAG87, KL2000, SECCHI2013221}. Thus, we provide a fully consistent framework to describe both equilibrium and dynamical magnetic properties of strongly interacting fermionic systems.

\section{Effective bosonic action for charge and spin degrees of freedom}
\label{sec:general}

In the current work we aim at deriving a quantum bosonic action that describes the behavior of the charge and spin degrees of freedom of an initially purely fermionic problem. 
We start with the lattice action of the extended Hubbard model written in the coordinate and imaginary time representation
\begin{gather}
{\cal S}_{\rm latt} = \int_{0}^{\beta} d\tau \, \Bigg\{  - \sum_{ij,\sigma\sigma'} c^{*}_{i\tau\sigma} \left[\delta_{ij}\delta_{\sigma\sigma'}(-\partial_{\tau}+\mu)-\varepsilon^{\sigma\sigma'}_{ij}\right]
c_{j\tau\sigma'}^{\phantom{*}} \notag\\
+ \sum_{i,\sigma\sigma'} U n_{i\tau\uparrow} n_{i\tau\downarrow} +\frac12 \sum_{ij,\varsigma} \rho^{\varsigma}_{i\tau} V^{\varsigma}_{ij} \, \rho^{\varsigma}_{j\tau} \Bigg\} 
\label{eq:action_latt}
\end{gather}
Fermionic Grassmann variables $c^{(*)}_{i\tau\sigma}$ describe annihilation (creation) of an electron with the spin projection ${\sigma=\{\uparrow, \downarrow\}}$ at the site $i$ and imaginary time $\tau$. 
$\mu$ is the chemical potential, and ${\varepsilon^{\sigma\sigma'}_{ij}}$ is the hopping matrix that has the following form in the spin space:
${\varepsilon^{\sigma\sigma'}_{ij} = \varepsilon_{ij}\,\delta_{\sigma\sigma'} + i\,\vec{\kappa}_{ij}\cdot\vec{\sigma}_{\sigma\sigma'}}$. 
The spin component-diagonal part $\varepsilon_{ij}$ of this matrix corresponds to the hopping amplitude of electrons between $i$ and $j$ lattice sites. 
The off-diagonal part $\vec{\kappa}_{ij}$ accounts for the spin-orbit coupling (SOC) in the Rashba form~\cite{Rashba, PhysRevB.52.10239}, where $\vec{\sigma}=\{\sigma^{x}, \sigma^{y}, \sigma^{z}\}$ is a vector of Pauli matrices.
$U$ is the on-site Coulomb repulsion and $V^{\varsigma}_{ij}$ describes the non-local (${V^{\varsigma}_{ii}=0}$) interaction between charge (${\varsigma=c}$) and spin (${\varsigma=s=\{x,y,z\}}$) densities ${n^{\varsigma}_{i\tau} = \sum_{\sigma\sigma'} c^{*}_{i\tau\sigma} \, \sigma^{\varsigma}_{\sigma\sigma'} c^{\phantom{*}}_{i\tau\sigma'}}$.
For convenience we introduce the variables ${\rho^{\varsigma}_{i\tau} = n^{\varsigma}_{i\tau} - \av{n^{\varsigma}}}$ that describe fluctuations of the densities around their average value.
We assume that the average densities can be obtained from a certain local reference system.

In this work the role of the reference system is played by an effective local site-independent impurity problem of the dynamical mean-field theory (DMFT)~\cite{RevModPhys.68.13}
\begin{gather}
{\cal S}_{\rm imp} = - \iint_{0}^{\beta} d\tau\,d\tau'
\sum_{\sigma\sigma'} c^{*}_{\tau\sigma} \left[\delta^{\phantom{*}}_{\tau\tau'}\delta^{\phantom{*}}_{\sigma\sigma'}(-\partial_{\tau}+\mu) - \Delta^{\sigma\sigma'}_{\tau\tau'}\right] c^{\phantom{*}}_{\tau'\sigma'} \notag\\
+\int_{0}^{\beta} d\tau \, U n_{\tau\uparrow} n_{\tau\downarrow}
\label{eq:action_imp}
\end{gather}
The advantage of considering this reference system is that it can be solved numerically exactly, e.g. by means of the continuous-time quantum Monte Carlo method~\cite{PhysRevB.72.035122, PhysRevLett.97.076405, PhysRevLett.104.146401, RevModPhys.83.349}.
Therefore, introducing such reference system makes investigation of local correlation effects more accessible.
In particular, this will help us to address the problem of the local moment formation in the system.
In order to isolate the impurity problem from the initial action~\eqref{eq:action_latt}, we add the fermionic hybridization function ${\Delta^{\sigma\sigma'}_{\tau\tau'} = \Delta^{\sigma\sigma'}(\tau-\tau')}$ to the local part of the lattice problem. 
To be consistent, the same hybridization is subtracted from the remaining (non-local) part of the lattice problem ${{\cal S}_{\rm rem} = {\cal S}_{\rm latt} - \sum_{i}{\cal S}_{\rm imp}}$.
This way of introducing the reference system gives some freedom in choosing the form of the hybridization function~\cite{BRENER2020168310}.
For instance, $\Delta_{\tau\tau'}$ does not necessarily have to be obtained from the DMFT self-consistency condition, which equates the local part of the lattice Green's function $G^{\sigma\sigma'}_{ii,\tau\tau'}$ to the exact local impurity Green's function $g^{\sigma\sigma'}_{\tau\tau'}$~\cite{RevModPhys.68.13}.
In this work we stick to the paramagnetic case, which is the most challenging regime for describing the behavior of the local magnetic moment.
Indeed, in the ordered state the value of the magnetic moment is given by the average magnetization, which in many cases can be obtained from the density-functional theory in reasonable agreement with the experiment~\cite{M3,Kubler}.
On the contrary, in the paramagnetic regime the average magnetization is equal to zero even if the magnetic moment has already been formed.
In the latter case the average magnetization is zero as a consequence of an uncorrelated precession of the magnetic moment, and distinguishing this situation from the case when the system does not possess any magnetic moment at all is a non-trivial task. 
As has been mentioned above, in the current work the average magnetization is given by the local reference system~\eqref{eq:action_imp}.
For this reason, we consider a spin-independent hybridization function ${\Delta^{\sigma\sigma'}_{\tau\tau'}=\delta^{\phantom{*}}_{\sigma\sigma'}\Delta^{\phantom{*}}_{\tau\tau'}}$, which ensures that the average local spin density is zero ${\av{n^{s}}_{\rm imp}=0}$, and therefore ${\rho^{s}_{i\tau} = n^{s}_{i\tau}}$.
As a consequence, the Green's function ${g^{\sigma\sigma'}_{\tau\tau'}=\delta^{\phantom{*}}_{\sigma\sigma'}g^{\phantom{*}}_{\tau\tau'}}$ of such reference system is also diagonal in the spin space. 
At the same time, the lattice Green's function can have non-diagonal spin components due to the presence of the SOC.
For this reason, we determine the hybridization function from the following self-consistency condition on the diagonal part of the lattice Green's function ${\frac12\sum_{\sigma}G^{\sigma\sigma}_{ii,\tau\tau'}=g^{\phantom{*}}_{\tau\tau'}}$.

We point out that $\rho^{\varsigma}$ is not a suitable variable for addressing the problem of charge and spin dynamics. Indeed, it is not a true bosonic field, because it is composed of two fermionic Grassmann variables.  
Proper bosonic variables that describe fluctuations of charge and spin densities can be introduced performing a set of Hubbard-Stratonovich transformations as has been shown in Refs.~\onlinecite{PhysRevLett.121.037204, PhysRevB.99.115124}.
Following the idea of these works we first rewrite the non-local part of the lattice action ${\cal S}_{\rm rem}$ in terms of new fermionic $f^{(*)}$ and truly bosonic $\phi^{\varsigma}$ fields instead of original fermionic $c^{(*)}$ and composite $\rho^{\varsigma}$ variables.
This transformation is explicitly shown in Appendix~\ref{app:HS} for a general multi-orbital case and results in the following lattice action
\begin{align} 
{\cal S}
= 
&-\int_{0}^{\beta} \{d\tau_{i}\} \sum_{ij,\sigma\sigma'} f^{*}_{i\tau_1\sigma}g^{-1}_{\tau_1\tau_2}
[\tilde{\varepsilon}^{-1}]^{\sigma\sigma'}_{ij,\tau_2\tau_3} \,
g^{-1}_{\tau_3\tau_4}f^{\phantom{*}}_{j\tau_4\sigma'}
+ \sum_{i}{\cal S}_{\rm imp} \notag\\
&-\frac12\int_{0}^{\beta} d\tau \sum_{ij,\varsigma}\phi^{\varsigma}_{i\tau} \left[V^{\varsigma}\right]^{-1}_{ij} \phi^{\varsigma}_{j\tau} 
+\int_{0}^{\beta} d\tau \sum_{i,\varsigma}
\left(\phi^{\varsigma}_{i\tau} + j^{\,\varsigma}_{i\tau}\right) \rho^{\varsigma}_{i\tau}\notag\\
&+\iint_{0}^{\beta} d\tau \, d\tau' \sum_{i,\sigma} \Big( c^{*}_{i\tau\sigma}g^{-1}_{\tau\tau'} 
f^{\phantom{*}}_{i\tau'\sigma} 
+ 
f^{*}_{i\tau\sigma} g^{-1}_{\tau\tau'} c^{\phantom{*}}_{i\tau'\sigma} \Big)
\label{eq:S1}
\end{align}
where $\tilde{\varepsilon}^{\sigma\sigma'}_{ij\tau\tau'} = \delta^{\phantom{*}}_{\tau\tau'}\varepsilon^{\sigma\sigma'}_{ij} - \delta^{\phantom{*}}_{ij}\delta^{\phantom{*}}_{\sigma\sigma'}\Delta^{\phantom{*}}_{\tau\tau'}$.
In Eq.~\eqref{eq:S1} we have introduced a source field $j^{\,\varsigma}$ for a composite $\rho^{\varsigma}$ variable. 
This source field will help us to identify the correct variables for original charge and spin degrees of freedom after multiple transformations of the initial action. 

Static properties of effective bosonic models for spin or charge degrees of freedom, namely the exchange interaction between spin or charge densities, have been studied in previous works~\cite{PhysRevLett.121.037204, PhysRevB.99.115124}.
Description of the spin dynamics, which has not been performed there, is a non-trivial task that requires a careful separation of the precession of the vector spin field from the fluctuation of the absolute value of the local magnetic moment. 
To this effect, we deviate from the main route of these works and make a transformation to a rotating frame for original fermionic variables
${c^{*}_{i\tau} \to c^{*}_{i\tau}R^{\phantom{*}}_{i\tau}}$ and ${c^{\phantom{*}}_{i\tau} \to R^{\dagger}_{i\tau} c^{\phantom{*}}_{i\tau}}$
introducing a unitary
matrix in the spin space
\begin{align}
R_{i\tau} = 
\begin{pmatrix}
\cos(\theta_{i\tau}/2) & - e^{-i\varphi_{i\tau}}\sin(\theta_{i\tau}/2) \\
e^{i\varphi_{i\tau}}\sin(\theta_{i\tau}/2) & \cos(\theta_{i\tau}/2)
\end{pmatrix}
\label{eq:Rmatrix}
\end{align}
where ${c_{i\tau} = (c_{i\tau\uparrow}, c_{i\tau\downarrow})^{T}}$. As is usually done in other works~\cite{PhysRevLett.65.2462, PhysRevB.43.3790, doi:10.1142/S0217979200002430, DUPUIS2001617}, we introduce additional functional integration over the rotation angles $\Omega_{R} = \{\theta_{i\tau}, \varphi_{i\tau}\}$.
Later on, we will associate these angles with the direction of the local magnetic moment and discuss when such an approximation is permitted.
Therefore, at each lattice site $i$ and imaginary time $\tau$ the rotation matrices are intended to adjust the coordinate system such that the local magnetic moment in new coordinates always points in $z$ direction.  
In this way the accounting for the rotation dynamics of the local magnetic moment is transferred from the corresponding bosonic field to a new time- and position-dependent coordinate system. 
Under this rotation the impurity problem transforms as
\begin{align}
{\cal S}_{\rm imp} &\to {\cal S}_{\rm imp} + 
\int_{0}^{\beta} d\tau \,\Tr_{\sigma} c^{*}_{i\tau} R^{\dagger}_{i\tau} \dot{R}^{\phantom{*}}_{i\tau} c^{\phantom{*}}_{i\tau} \notag\\
&\,=\, {\cal S}_{\rm imp} +
\int_{0}^{\beta} d\tau \sum_{\varsigma} {\cal A}^{\varsigma}_{i\tau} \, \rho^{\varsigma}_{i\tau}
\end{align}
where $\dot{R}^{\phantom{*}}_{i\tau} = \partial_{\tau} R^{\phantom{*}}_{i\tau}$ and ${\cal A}^{\varsigma}_{i\tau}$ is an effective gauge field introduced as ${R^{\dagger}_{i\tau} \dot{R}^{\phantom{*}}_{i\tau} = \sum_{\varsigma} {\cal A}^{\varsigma}_{i\tau}\sigma^{\varsigma}}$.
The explicit form of the rotation matrix~\eqref{eq:Rmatrix} implies that ${\cal A}^{c}_{i\tau}=0$.
Composite variables for charge and spin degrees of freedom become
\begin{align}
\rho^{\varsigma}_{i\tau} \to \sum_{\varsigma'} {\cal U}^{\varsigma\varsigma'}_{i\tau}\rho^{\varsigma'}_{i\tau}
\end{align}
where ${\cal U}^{\varsigma\varsigma'}_{i\tau}$ satisfies
\begin{align}
R^{\dagger}_{i\tau} \sigma^{\varsigma} R^{\phantom{*}}_{i\tau} 
= \sum_{\varsigma'} {\cal U}^{\varsigma\varsigma'}_{i\tau}\sigma^{\varsigma'}
\label{eq:B_coupling}
\end{align}
It can be shown that ${\cal U}^{ss'}$ is a unitary matrix, i.e. ${[{\cal U}^{-1}]^{ss'}=[{\cal U}^{\rm T}]^{ss'}}$, and that ${{\cal U}^{cs}_{i\tau}=0}$ and ${{\cal U}^{cc}_{i\tau}=1}$.
The last equality originates from the fact that the charge density $n^{c}_{i\tau}$ is invariant under rotation in the spin space.
Upon collecting all terms, the lattice action~\eqref{eq:S1} transforms to
\begin{align} 
{\cal S} = 
&-\int_{0}^{\beta} \{d\tau_{i}\} \sum_{ij,\sigma\sigma'} f^{*}_{i\tau_1\sigma}g^{-1}_{\tau_1\tau_2}
[\tilde{\varepsilon}^{-1}]^{\sigma\sigma'}_{ij,\tau_2\tau_3} \,
g^{-1}_{\tau_3\tau_4}f^{\phantom{*}}_{j\tau_4\sigma'} \notag\\
&-\frac12\int_{0}^{\beta} d\tau \sum_{ij,\varsigma}\phi^{\varsigma}_{i\tau} \left[V^{\varsigma}\right]^{-1}_{ij} \phi^{\varsigma}_{j\tau}
+ \sum_{i}{\cal S}_{\rm imp} \notag\\
&+ \iint_{0}^{\beta} d\tau \, d\tau' \,\Tr_{\sigma}\sum_{i} \Bigg\{ c^{*}_{i\tau} R^{\dagger}_{i\tau} g^{-1}_{\tau\tau'}
f^{\phantom{*}}_{i\tau'}
+ f^{*}_{i\tau} g^{-1}_{\tau\tau'}
R^{\phantom{\dagger}}_{i\tau'} c^{\phantom{*}}_{i\tau'} \Bigg\} \notag\\
&+ \int_{0}^{\beta} d\tau \,\sum_{i,\varsigma\varsigma'}\left(\phi^{\varsigma}_{i\tau} + j^{\,\varsigma}_{i\tau} + \sum_{\varsigma''} {\cal A}^{\varsigma''}_{i\tau} [{\cal U}_{i\tau}^{-1}]^{\varsigma''\varsigma}\right) {\cal U}^{\varsigma\varsigma'}_{i\tau}\rho^{\varsigma'}_{i\tau} 
\label{eq:S2}
\end{align}
We find that the bosonic field $\phi^{s}_{i\tau}$ enters the lattice action~\eqref{eq:S2} as an effective magnetic field.
However, it is important to emphasise that the local magnetic moment at a given lattice site does not necessarily point in the same direction as the polarizing field applied to the same site. Therefore we prefer to keep this freedom and do not relate the polar angles $\theta_{i\tau}$ and $\varphi_{i\tau}$ with the direction of the field $\phi^{s}_{i\tau}$ as commonly used at similar transformations of the effective action (see e.g.~\cite{sachdev2008quantum, PhysRevX.8.011012, ScheurerE3665, PhysRevX.8.021048, PhysRevX.10.041057}).
Instead, below we demonstrate an alternative way of introducing the bosonic field that describes Higgs fluctuations of the local magnetic moment rather than of its conjugated field.

After the rotational dynamics of the magnetic moment is explicitly isolated, original fermionic variables can be integrated out.
This allows one to account for local correlation effects exactly via the reference system~\eqref{eq:action_imp}, which is formulated solely in terms of original variables. 
It is important that upon all transformations of the lattice problem the source field $j^{\,\varsigma}$ and the effective gauge field ${\cal A}^{\varsigma}$ be taken into account exactly without any approximation.
For this purpose we make the following shift of variables
\begin{align}
\phi^{\varsigma}_{i\tau} &\to \hat{\phi}^{\varsigma}_{i\tau} = \phi^{\varsigma}_{i\tau} - j^{\,\varsigma}_{i\tau} - \sum_{\varsigma''}{\cal A}^{\varsigma''}_{i\tau} [{\cal U}_{i\tau}^{-1}]^{\varsigma''\varsigma}
\end{align}
that excludes $j^{\,\varsigma}$ and ${\cal A}^{\varsigma}$ fields from the integration of original fermionic degrees of freedom. 
When the latter are integrated out, we transform auxiliary bosonic fields $\hat{\phi}^{\varsigma}$ to physical variables $\bar{\rho}^{\,\varsigma}$ that describe fluctuations of charge and spin densities.
For a general multi-orbital case all these steps are performed in details in Appendix~\ref{app:HS}, which results in the effective fermion-boson action
\begin{align} 
{\cal S}
= 
&-\int_{0}^{\beta} \{d\tau_{i}\} \sum_{ij,\sigma\sigma'} f^{*}_{i\tau_1\sigma}g^{-1}_{\tau_1\tau_2}
[\tilde{\varepsilon}^{-1}]^{\sigma\sigma'}_{ij,\tau_2\tau_3} \,
g^{-1}_{\tau_3\tau_4}f^{\phantom{*}}_{j\tau_4\sigma'} \notag\\
&+ \int_{0}^{\beta} \{d\tau_{i}\} \Tr_{\sigma} \sum_{i} f^{*}_{i\tau_1}g^{-1}_{\tau_1\tau_2} R^{\phantom{\dagger}}_{i\tau_2}g^{\phantom{*}}_{\tau_2\tau_3}R^{\dagger}_{i\tau_3}g^{-1}_{\tau_3\tau_4}f^{\phantom{*}}_{i\tau_4}  \notag\\
&+ \frac12\iint_{0}^{\beta} d\tau\,d\tau' \sum_{i,\{\varsigma\}} \bar{\rho}^{\,\varsigma_1}_{i\tau} {\cal U}^{\varsigma_1\varsigma_2}_{i\tau} \left[\chi^{\varsigma_2}\right]^{-1}_{\tau\tau'} [{\cal U}_{i\tau'}^{-1}]^{\varsigma_2\varsigma_3} \bar{\rho}^{\,\varsigma_3}_{i\tau'}  \notag\\ 
&+ \frac12 \int_{0}^{\beta} d\tau \sum_{ij,\varsigma} \bar{\rho}^{\,\varsigma}_{i\tau} V^{\varsigma}_{ij} \, \bar{\rho}^{\,\varsigma}_{j\tau} \notag\\
&+ \int_{0}^{\beta} \{d\tau_{i}\} \Tr_{\sigma} \sum_{i,\varsigma\varsigma'} f^{*}_{i\tau_1}
R^{\phantom{\dagger}}_{i\tau_1}\sigma^{\varsigma}R^{\dagger}_{i\tau_2}
f^{\phantom{*}}_{i\tau_2} \Lambda^{\varsigma}_{\tau_1\tau_2\tau_3} [{\cal U}_{i\tau_3}^{-1}]^{\varsigma\varsigma'} \bar{\rho}^{\,\varsigma'}_{i\tau_3} \notag\\
&+ \int_{0}^{\beta} d\tau \sum_{i} \Bigg\{ \sum_{ss'}{\cal A}^{s}_{i\tau}[{\cal U}_{i\tau}^{-1}]^{ss'} \bar{\rho}^{\,s'}_{i\tau}
+ \sum_{\varsigma} j^{\,\varsigma}_{i\tau}\,\bar{\rho}^{\,\varsigma}_{i\tau} \Bigg\}
\label{eq:S3}
\end{align}
Quantities $g_{\tau\tau'}$, $\chi^{\varsigma}_{\tau\tau'}$, and $\Lambda^{\varsigma}_{\tau_1\tau_2\tau_3}$ are respectively the exact Green's function, the susceptibility, and the three-point vertex function of the reference system~\eqref{eq:action_imp}.
They are explicitly defined in Appendix~\ref{app:HS}.
In analogy to $s\text{-}d$ exchange model~\cite{M2}, 
$\Lambda^{\varsigma}_{\tau_1\tau_2\tau_3}$ can be seen as a renormalized local coupling between fermions $f^{(*)}$ and charge or spin densities $\bar{\rho}^{\varsigma}$.

We find that the source field $j^{\,\varsigma}_{i\tau}$ enters the new lattice problem~\eqref{eq:S3} only multiplied by the new bosonic field $\bar{\rho}^{\,\varsigma}_{i\tau}$. 
This means that all correlation functions written in terms of original composite variables $\rho^{\varsigma}_{i\tau}$ identically coincide with the ones where the $\rho^{\varsigma}_{i\tau}$ variables are replaced by the corresponding bosonic fields $\bar{\rho}^{\,\varsigma}_{i\tau}$.
Therefore, the introduced fields $\bar{\rho}^{\,\varsigma}_{i\tau}$ have the same physical meaning as the composite variables $\rho^{\varsigma}_{i\tau}$ that describe fluctuations of charge and spin densities.
To emphasise this point hereinafter we omit the bar over the $\bar{\rho}^{\,\varsigma}_{i\tau}$ field.

The dynamics of the bosonic field $\rho^{\varsigma}$ is described by the 3rd and 6th terms of Eq.~\eqref{eq:S3}.
It is convenient to rewrite the vector bosonic field as ${\rho^{s}_{i\tau} = M^{\phantom{*}}_{i\tau} e^{s}_{i\tau}}$, where $M_{i\tau}$ is a scalar field that describes Higgs fluctuations of the absolute value of the local magnetic moment. In turn, $\vec{e}_{i\tau}$ is the unit vector field, defined by the angles $\Omega_M=\{\theta'_{i\tau}, \varphi'_{i\tau}\}$, that points in the direction of the local moment on the lattice site $i$ at the time $\tau$. It can be shown (for details see Appendix~\ref{app:SPA}) that under the adiabatic approximation, for parameters that allow formation of a sufficiently large well-defined local moment, the path integral over $\vec{e}_{i\tau}$ in the saddle approximation yields that the direction of the vector bosonic field can be pinned to the $z$-axis of the rotating reference frame. At this point it is worth mentioning that the local moment is likely to be large enough for the saddle approximation to be applicable only in the multi-orbital case. This means that the Berry-phase term and consequently the Landau-Lifshitz-Gilbert equations of motion that are derived below work only in that regime. If on the other hand the saddle approximation can not reliably be applied, the quantum fluctuations play a decisive role in the dynamics of the local moment, and one can no longer speak about local moment dynamics in terms of classical equations of motion. 

From this point on until the end of the section as well as for the next section, we consider the integral over $\vec{e}_{i\tau}$ taken and the two sets of rotation angles are set equal to one another $\Omega_{R}=\Omega_{M}$.
This results in the following relation for the unit vector field
\begin{align}
R^{\dagger}_{i\tau} \, \vec{\sigma} \cdot \vec{e}^{\phantom{*}}_{i\tau} \, R^{\phantom{*}}_{i\tau} 
= \sigma^{z}
\end{align}
Using the relation~\eqref{eq:B_coupling}, one also finds that 
\begin{align}
\sum_{s'} [{\cal U}_{i\tau}^{-1}]^{ss'} e^{s'}_{i\tau} =  \delta^{\phantom{*}}_{s,z} 
\end{align}
which immediately yields the Berry phase term
\begin{align}
\sum_{ss'}{\cal A}^{s}_{i\tau}[{\cal U}_{i\tau}^{-1}]^{ss'} \rho^{s'}_{i\tau} =
\sum_{ss'}{\cal A}^{s}_{i\tau}[{\cal U}_{i\tau}^{-1}]^{ss'} e^{s'}_{i\tau} M^{\phantom{*}}_{i\tau} = 
{\cal A}^{z}_{i\tau}M^{\phantom{*}}_{i\tau}
\end{align}
where ${{\cal A}^{z}_{i\tau} = \frac{i}{2} \dot{\varphi}_{i\tau} (1-\cos\theta_{i\tau})}$.
Also, one straightforwardly gets that
\begin{align}
\sum_{\{s\}}\rho^{\,s_1}_{i\tau} {\cal U}^{s_1s_2}_{i\tau} \left[\chi^{s_2}\right]^{-1}_{\tau\tau'} [{\cal U}_{i\tau'}^{-1}]^{s_2s_3} \rho^{\,s_3}_{i\tau'} = M^{\phantom{*}}_{i\tau} \left[\chi^{z}\right]^{-1}_{\tau\tau'} M^{\phantom{*}}_{i\tau'}
\end{align}
Remaining rotation matrices can be excluded from the action~\eqref{eq:S3} by assuming the adiabatic approximation that characteristic times for electronic degrees of freedom are much faster than for spin ones. In this framework $g^{\phantom{\dagger}}_{\tau_2\tau_3}$ changes much faster than the rotation matrices $R^{\phantom{\dagger}}_{i\tau_2}$. 
This approximation results in
\begin{gather}
\iint^{\beta}_{0} d\tau_2 \, d\tau_3 \, g^{-1}_{\tau_1\tau_2} R^{\phantom{\dagger}}_{i\tau_2} g^{\phantom{\dagger}}_{\tau_2\tau_3} R^{\dagger}_{i\tau_3} g^{-1}_{\tau_3\tau_4} \simeq \notag\\ 
\int^{\beta}_{0} d\tau_3 \, \delta^{\phantom{*}}_{\tau_1\tau_3} R^{\phantom{\dagger}}_{i\tau_3}R^{\dagger}_{i\tau_3}g^{-1}_{\tau_3\tau_4} 
= g^{-1}_{\tau_1\tau_4}
\end{gather}
because the local reference system~\eqref{eq:action_imp} is non-polarized, and its exact Greens function $g_{\tau\tau'}$ is diagonal in the spin space. 
A similar trick can be performed for the three-point vertex function, which leads to (see Appendix~\ref{app:HS})
\begin{gather}
\Tr_{\sigma}\sum_{ss'} f^{*}_{i\tau_1}
R^{\phantom{\dagger}}_{i\tau_1}\sigma^{s}R^{\dagger}_{i\tau_2}
f^{\phantom{*}}_{i\tau_2} \Lambda^{\hspace{-0.05cm}s}_{\tau_1\tau_2\tau_3} [{\cal U}_{i\tau_3}^{-1}]^{ss'} \rho^{s'}_{i\tau_3} 
\simeq \notag\\
\sum_{s,\sigma\sigma'} f^{*}_{i\tau_1\sigma}
\sigma^{s}_{\sigma\sigma'}
f^{\phantom{*}}_{i\tau_2\sigma'} \Lambda^{\hspace{-0.05cm}s}_{\tau_1\tau_2\tau_3}\,
\rho^{\,s}_{i\tau_3}
\end{gather}
The action~\eqref{eq:S3} becomes
\begin{align} 
{\cal S}
= 
&-\iint_{0}^{\beta} d\tau \, d\tau' \sum_{ij,\sigma\sigma'}  
f^{*}_{i\tau\sigma} \left[\tilde{\cal G}^{-1} \right]^{\sigma\sigma'}_{ij,\tau\tau'}
f^{\phantom{*}}_{j\tau'\sigma'} \notag\\
&+ \iiint_{0}^{\beta} d\tau_1 \, d\tau_2 \, d\tau_3 \sum_{i,\varsigma,\sigma\sigma'} f^{*}_{i\tau_1\sigma}\sigma^{\varsigma}_{\sigma\sigma'}
f^{\phantom{*}}_{i\tau_2\sigma'} \,\Lambda^{\varsigma}_{\tau_1\tau_2\tau_3}\, \rho^{\varsigma}_{i\tau_3} \notag\\
&+ \frac12 \int_{0}^{\beta} d\tau\,\sum_{ij,\varsigma} \rho^{\varsigma}_{i\tau} V^{\varsigma}_{ij} \, \rho^{\varsigma}_{j\tau}
+ \int_{0}^{\beta} d\tau \sum_{i} {\cal A}^{z}_{i\tau} M^{\phantom{*}}_{i\tau} \notag\\
& - \frac12 \iint_{0}^{\beta} d\tau\,d\tau' \sum_{i} \Bigg\{ \rho^{c}_{i\tau} \left[\chi^{c}\right]^{-1}_{\tau\tau'} \rho^{c}_{i\tau'} 
+ M^{\phantom{*}}_{i\tau} \left[\chi^{z}\right]^{-1}_{\tau\tau'} M^{\phantom{*}}_{i\tau'} \Bigg\} 
\label{eq:S4}
\end{align}
Source fields $j^{\,\varsigma}$ have been introduced only to identify correct bosonic variables and were excluded from the action~\eqref{eq:S4}.
The bare Green's function of the action~\eqref{eq:S4} is given by the difference between the DMFT and the impurity Green's functions~\cite{PhysRevB.77.033101, PhysRevB.79.045133}
\begin{align}
\tilde{\cal G}^{\sigma\sigma'}_{ij,\tau\tau'} = G^{\sigma\sigma'}_{ij,\tau\tau'} - \delta^{\phantom{*}}_{ij}\delta^{\phantom{*}}_{\sigma\sigma'} g^{\phantom{*}}_{\tau\tau'}
\label{eq:G_dual}
\end{align}
Therefore, this quantity is dressed only in the local impurity self-energy, which in our case is obtained from the self-consistent DMFT calculation.
For this reason, the developed formalism does not suffer from a causality problem~\cite{RevModPhys.68.13, 2020arXiv201105311B}.
In the absence of the SOC the local part of the bare dual Green's function is identically zero due to DMFT self-consistency condition.
Therefore, the performed set of transformations separates spatial electronic fluctuations described by fields $f^{(*)}$ from local correlation effects that are accounted for by the reference system~\eqref{eq:action_imp}.
Another advantage of the introduced fermion boson action~\eqref{eq:S4} is that it allows one to obtain not only standard correlation functions, such as the electronic Green's function and the (charge, spin, etc.) susceptibility, but also various exchange interactions between charge and spin densities that cannot be calculated directly from the initial electronic problem~\eqref{eq:action_latt}.
To illustrate this point one can integrate out fermionic fields to get an effective bosonic action
\begin{align} 
{\cal S}
= 
& - \Tr\ln \left[ \left[\tilde{\cal G}^{-1} \right]^{\sigma\sigma'}_{ij,\tau\tau'} - \delta^{\phantom{*}}_{ij} \int^{\beta}_{0} d\tau'' \sum_{\varsigma} \, \sigma^{\varsigma}_{\sigma\sigma'} \Lambda^{\varsigma}_{\tau\tau'\tau''} \, \rho^{\varsigma}_{i\tau''} \right]
\notag\\
&+ \frac12 \int_{0}^{\beta} d\tau\,\sum_{ij,\varsigma} \rho^{\varsigma}_{i\tau} V^{\varsigma}_{ij} \, \rho^{\varsigma}_{j\tau}
+ \int_{0}^{\beta} d\tau \sum_{i} {\cal A}^{z}_{i\tau} M^{\phantom{*}}_{i\tau} \notag\\
& - \frac12 \iint_{0}^{\beta} d\tau\,d\tau' \sum_{i} \Bigg\{ \rho^{c}_{i\tau} \left[\chi^{c}\right]^{-1}_{\tau\tau'} \rho^{c}_{i\tau'} 
+ M^{\phantom{*}}_{i\tau} \left[\chi^{z}\right]^{-1}_{\tau\tau'} M^{\phantom{*}}_{i\tau'} \Bigg\} 
\label{eq:S5}
\end{align}
The first line in this expression describes all possible (non-local) exchange interactions that can be obtained expanding this part of equation in terms of the $\rho^{\varsigma}$ variables.
The first order contribution in this expansion results in an effective local magnetic field 
\begin{align}
h^{{\rm soc}\,s}_{i\tau_3} = -\iint_{0}^{\beta} d\tau_1 \, d\tau_2  \sum_{\sigma\sigma'} \tilde{\cal G}^{\sigma'\sigma}_{ii,\tau_2\tau_1} \sigma^{s}_{\sigma\sigma'} \Lambda^{s}_{\tau_1\tau_2\tau_3}
\label{eq:h_soc}
\end{align}
which is identically zero in the absence of the SOC due to non-locality of the bare Green's function~\eqref{eq:G_dual}, namely ${\tilde{\cal G}^{\sigma'\sigma}_{ii,\tau_2\tau_1}=0}$. 
Since the SOC enters the problem~\eqref{eq:action_latt} as a non-local hopping ${\vec{\kappa}_{ij}\cdot\vec{\sigma}_{\sigma\sigma'}}$, the local Green's function $\tilde{\cal G}^{\sigma'\sigma}_{ii,\tau_2\tau_1}$ is also negligibly small in the case of a small SOC.
Quadratic exchange can be obtained as the second order of the expansion, which gives
\begin{align} 
{\cal S}
&\simeq 
\frac12 \iint_{0}^{\beta} d\tau\,d\tau'\sum_{ij,\varsigma\varsigma'} \rho^{\varsigma}_{i\tau} {\cal I}^{\varsigma\varsigma'}_{ij,\tau\tau'} \, \rho^{\varsigma'}_{j\tau'} 
+ \int_{0}^{\beta} d\tau \sum_{i} {\cal A}^{z}_{i\tau} M^{\phantom{*}}_{i\tau} \notag\\
&\,- \frac12 \iint_{0}^{\beta} d\tau\,d\tau' \sum_{i} \Bigg\{ \rho^{c}_{i\tau} \left[\chi^{c}\right]^{-1}_{\tau\tau'} \rho^{c}_{i\tau'} 
+ M^{\phantom{*}}_{i\tau} \left[\chi^{z}\right]^{-1}_{\tau\tau'} M^{\phantom{*}}_{i\tau'} \Bigg\} \notag\\
&-\int_{0}^{\beta} d\tau \sum_{i,s} h^{{\rm soc}\,s}_{i\tau} \rho^{s}_{i\tau}
\label{eq:S6}
\end{align}
The total non-local quadratic exchange interaction
\begin{align}
{\cal I}^{\varsigma\varsigma'}_{ij,\tau\tau'} = \delta^{\phantom{*}}_{\tau\tau'} \delta^{\phantom{*}}_{\varsigma\varsigma'} V^{\varsigma}_{ij} + J^{\varsigma\varsigma'}_{ij,\tau\tau'}
\label{eq:exch}
\end{align}
contains the bare (direct) interaction $V^{\varsigma}_{ij}$ of the initial action~\eqref{eq:action_latt} and the RKKY-type (kinetic) interaction mediated by electrons
\begin{align}
J^{\varsigma\varsigma'}_{ij,\tau\tau'} = 
\int_{0}^{\beta} \{d\tau_i\} \sum_{\{\sigma_i\}} \Lambda^{\hspace{-0.05cm}*\,\varsigma}_{\tau\tau_1\tau_2} \tilde{\cal G}^{\sigma_1\sigma_3}_{ij,\tau_1\tau_3} \tilde{\cal G}^{\sigma_4\sigma_2}_{ji,\tau_4\tau_2} \Lambda^{\hspace{-0.05cm}\varsigma'}_{\tau_3\tau_4\tau'}
\label{eq:J}
\end{align}
Here, the ``transposed'' three-point vertex ${\Lambda^{\hspace{-0.05cm}*\,\varsigma}_{\tau_1\tau_2\tau_3} = \Lambda^{\varsigma}_{\tau_3\tau_2\tau_1}}$ is introduced to simplify notations.
The diagonal part of the kinetic interaction is given by the Heisenberg exchange interaction $J^{ss}_{ij}$ for spin~\cite{PhysRevLett.121.037204} and the Ising interaction $J^{cc}_{ij}$ for charge~\cite{PhysRevB.99.115124} densities.
The non-diagonal $J^{ss'}_{ij}$ (${s\neq{}s'}$) component appears due to the SOC and gives rise to the antisymmetric anisotropic (Dzyaloshinskii-Moriya) and the symmetric anisotropic interactions (see, e.g., Ref.~\onlinecite{PhysRevB.52.10239}).
More involved interaction terms, as for example the chiral three-spin~\cite{PhysRevLett.93.056402, bauer2014chiral, PhysRevB.95.014422, grytsiuk2020topological, zhang2020imprinting, PhysRevB.103.L060404} 
and the four-spin~\cite{PhysRevB.76.054427,heinze2011spontaneous, paul2020role} exchange interactions can be obtained by a straightforward expansion of the first term in Eq.~\eqref{eq:S5} to higher orders in $\rho$.
We note that the exchange interaction~\eqref{eq:exch} can also be seen as the non-local part of the inverse of the lattice susceptibility~\cite{PhysRevLett.121.037204, PhysRevB.99.115124}.
This fact clarifies the relation between our result~\eqref{eq:exch} and the estimation for the exchange interaction based on the DMFT approximation for the susceptibility introduced previously~\cite{PhysRevB.91.195123, PhysRevB.96.075108, PhysRevB.99.165134}.

\begin{figure}[t!]
\includegraphics[width=0.8\linewidth]{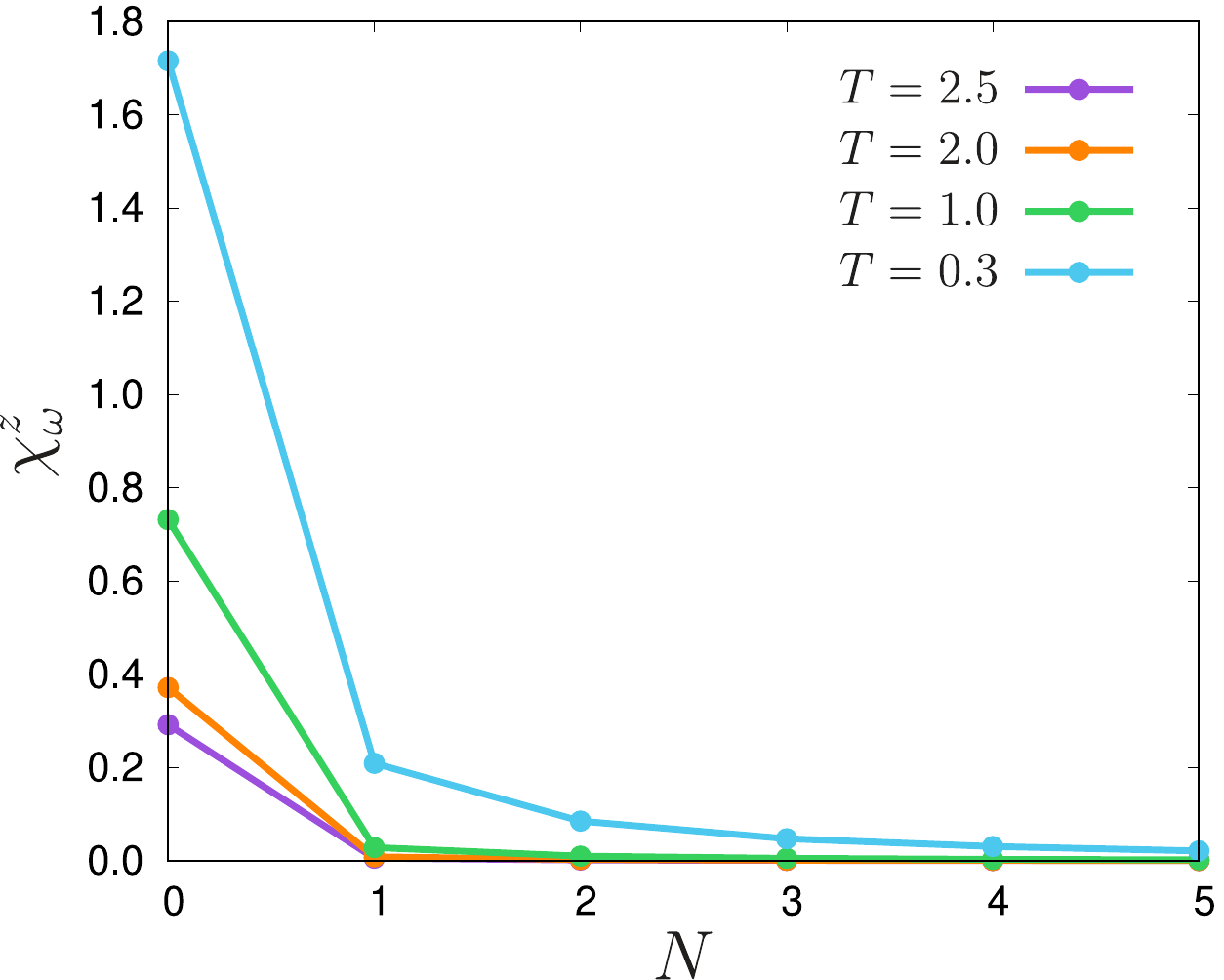}
\caption{\label{fig:chi_loc} Absolute value of the local spin susceptibility $\chi^{z}_{\omega}$ of the reference DMFT impurity problem~\eqref{eq:action_imp} as a function of the number $N$ of bosonic Matsubara frequency $\omega_{N}=2\pi{}N/\beta$. Results are obtained for $U=8$ at different temperatures specified in the legend. For all considered temperatures the local spin susceptibility shows a rapidly decaying behavior in the frequency space.}
\end{figure}

\section{Equation of motion}

In this section we derive equation of motion for the precession of the local magnetic moment and thus exclude charge degrees of freedom from consideration. 
The last term in the second line of Eq.~\eqref{eq:S6} describes the Higgs dynamics of the absolute value of the local magnetic moment $M_{i\tau}$.
These fluctuations are fast and the corresponding contribution is strongly non-local in time. 
This fact is confirmed by a rapidly decaying behavior of the Fourier transform of the local spin susceptibility $\chi^{s}_{\omega}$ to the Matsubara frequency space $\omega$ shown in Fig.~\ref{fig:chi_loc}. 
On the contrary, the precession of the local magnetic moment is slow in time~\cite{Sayad_2015, PhysRevLett.117.127201, PhysRevLett.125.086402} and can be described by the Landau-Lifshitz-Gilbert equation of motion~\cite{M200,M300}. 
To derive this equation we assume that the local magnetic moment has already been formed in the system. 
The criterion for the formation of the magnetic moment is discussed in details in the Section~\ref{sec:local_moment}.
At this point we average over fast Higgs fluctuations and replace the scalar field $M_{i\tau}$ by its constant non-zero average value ${\langle M_{i\tau} \rangle = 2S}$.
In this case the Higgs term can be neglected in the action, because now it only gives a constant contribution to the energy.
The bosonic action~\eqref{eq:S6} reduces to an effective spin problem  
\begin{align}
{\cal S}_{\rm spin} = \int^{\beta}_0 d\tau\sum_{j} \left( i\dot{\varphi}_{j\tau}(1-\cos\theta_{j\tau})\,S - \vec{S}_{\hspace{-0.05cm}j\tau}\cdot\vec{h}_{j\tau} \right)
\label{eq:action_spin}
\end{align}
where ${\vec{S}_{\hspace{-0.05cm}i\tau} = S\vec{e}_{i\tau}}$, and components of an effective magnetic field $\vec{h}_{j\tau}$ are following 
\begin{align}
h^{s}_{j\tau} = -~4 \int_0^{\beta} d\tau' \sum_{i,s'}\mathcal{I}^{ss'}_{ji,\tau\tau'}S^{\hspace{-0.05cm}s'}_{\hspace{-0.05cm}i\tau'}
+ h^{{\rm soc}\,s}_{j\tau}
\label{eq:h_eff}
\end{align}

In the general case the equation of motion for the spins is a set of integro-differential equations. 
To simplify the problem, we make use of the fact that the interaction between spins is determined by the super-exchange processes due to electrons~\eqref{eq:J} and thus decays fast on the time scales of inverse band width, while the time-dependence of the angle variables $\varphi_{i\tau}$ and $\theta_{i\tau}$ is slow.
For this reason, we can expand the time-dependence of the spin variable $S^{\hspace{-0.05cm}s'}_{\hspace{-0.05cm}i\tau'}$ in Eq.~\eqref{eq:h_eff} up to the first order in powers of ${\tau-\tau'}$.
In the zeroth order the $\tau'$ time argument of $S^{\hspace{-0.05cm}s'}_{\hspace{-0.05cm}i\tau'}$ is simply replaced by $\tau$.
Then, the $\tau'$ integration of ${{\cal I}^{ss'}_{ji}(\tau-\tau')}$ leads to the zero Matsubara frequency Fourier component of the spin-spin interaction ${{\cal I}^{ss'}_{ij}(\omega=0)}$, and the zeroth-order contribution to the effective magnetic field~\eqref{eq:h_eff} becomes 
\begin{align}
h^{0s}_{j\tau} = -~4\sum_{i,s'} {\cal I}^{ss'}_{ji,\omega=0} \, S^{\hspace{-0.05cm}s'}_{\hspace{-0.05cm}i\tau}
\end{align}
We note that the local three-point vertex function that enters the expression for the kinetic interaction $J^{ss'}_{ij,\omega=0}$~\eqref{eq:J} can be obtained from the self-energy of the impurity problem as~\cite{PhysRevLett.121.037204}
\begin{align}
\Lambda^{\hspace{-0.05cm}s}_{\nu,\omega=0} = \frac{\partial{\Sigma^{\rm imp}_{\nu}}}{\partial{}M_{\hspace{-0.05cm}\omega=0}} + \left(\chi^{s}_{\omega=0}\right)^{-1}
\label{eq:vertex_sigma}
\end{align}
where $\nu$ is the fermionic Matsubara frequency.
If the inverse of the local susceptibility is neglected, the equal-time kinetic interaction reduces to well-known expression
\begin{align}
J^{ss'}_{ij,\omega=0} = 
\sum_{\nu,\{\sigma\}} \left(\partial_{M}\Sigma^{s}_{i\nu}\right) \tilde{\cal G}^{\sigma_1\sigma_3}_{ij,\nu}
\left(\partial_{M}\Sigma^{s'}_{j\nu}\right)
\tilde{\cal G}^{\sigma_4\sigma_2}_{ji,\nu}
\label{eq:J_w0}
\end{align}
that has been derived in Refs.~\onlinecite{LKG84, LKG85, LKAG87, KL2000} for the exchange interaction in a magnetically ordered state based on magnetic force theorem. 
For a paramagnetic case this form for the exchange interaction has been obtained in Ref.~\onlinecite{PhysRevB.94.115117} using Hubbard-I approximation.
Within this approximation the local reference system~\eqref{eq:action_imp} is considered in the atomic limit that corresponds to a zero fermionic hybridization function $\Delta=0$.
In the presence of the SOC the simplified form~\eqref{eq:J_w0} of the effective exchange~\eqref{eq:J} gives a result for the Dzyaloshinskii-Moriya and the symmetric anisotropic interactions similar to the ones derived in Refs.~\onlinecite{KL2000, 2020arXiv200304680C} using the magnetic force theorem. 

The first order expansion of $S^{\hspace{-0.05cm}s'}_{\hspace{-0.05cm}i\tau'}$ in ${\tau-\tau'}$ in Eq.~\eqref{eq:h_eff} would lead to a term proportional to the time derivative of the spin in the effective field~\eqref{eq:h_eff}, which would be the Gilbert damping. 
But this term vanishes in the imaginary time due to the exchange ${{\cal I}^{ss'}_{ji}(\tau-\tau')}$ being an even function of time. 
This is not really surprising as one can not expect dissipation effects to be visible in the equilibrium formalism. 
On the other hand, if at this point we perform analytical continuation to real times $t$, the exchange transforms to a retarded function $I^{{\rm R}\,ss'}_{ji}(t-t')$, and the contribution with the first order time derivative of the spin in Eq.~\eqref{eq:h_eff} does not vanish. 
Up to this order we can write
\begin{align}
\vec{h}_{j}(t) = \vec{h}^{\rm soc}_{j}(t) + \vec{h}^{\,0}_{j}(t) - 4\sum_{i}\left.\left(\partial_{\Omega} \, {\rm Im}\, I^{{\rm R}\,ss}_{ji,\Omega}\right)\right|_{\Omega=0} \, \dot{\vec{S}}_{\hspace{-0.05cm}i}(t)
\label{eq:h_eff_fin}
\end{align}
where the second term describes the Gilbert damping for which we keep only the leading diagonal (${s=s'}$) component of the exchange interaction.
$I^{{\rm R}\,ss'}_{ji}(\Omega)$ is a Fourier transform of the retarded exchange interaction $I^{{\rm R}\,ss'}_{ji}(t-t')$ to real frequency $\Omega$.
We note that a similar expression for the Gilbert damping has been derived in Refs.~\onlinecite{Sayad_2015, PhysRevLett.117.127201} for the case of a classical spin coupled to the system of conduction electrons.

The derived effective magnetic field~\eqref{eq:h_eff_fin} allows one to obtain the equation of motion for the spin precession in the conventional Landau-Lifshitz-Gilbert form directly varying the action~\eqref{eq:action_spin}
\begin{align}
\dot{\vec{S}}_{\hspace{-0.05cm}j}(t) = -\, \vec{h}_{j}(t)\times\vec{S}_{\hspace{-0.05cm}j}(t)
\label{eq:eq_motion}
\end{align}
A more general expression for the equation of motion that accounts for all components of the exchange interaction can be found in Appendix~\ref{app:eq_motion}.

At this point we should emphasize again that our aim was a mapping of the initial interacting fermionic problem onto and effective Hamiltonian problem~\eqref{eq:action_spin} that is stationary in time. This effective problem describes the dynamics of spin degrees of freedom, which is supposed to be much slower than the electron hopping and other fast electron processes, in particular, related to the Hubbard $U$ energy scale. 
Within this approach, we should take into account only low-frequency part of the exchange term~\eqref{eq:J}, which is approximately limited by the value of the exchange interaction. 
Actually, the exchange term~\eqref{eq:J} has a complicated frequency dependence, in fact it diverges for high frequencies, but taking into account such non-adiabatic effects is not allowed in the derived Landau-Lifshitz-Gilbert equation of motion~\eqref{eq:eq_motion}. 
In the high frequency region the separation of spin and electron dynamics is, generally speaking, impossible.
In the latter case, the dynamics of charge and spin degrees of freedom can only by described by the derived fermion-boson~\eqref{eq:S4} or boson~\eqref{eq:S5} actions that have no restriction on the regime of frequencies, but are non-stationary in time.

\section{Local magnetic moment formation}
\label{sec:local_moment}

The introduced equation of motion~\eqref{eq:eq_motion} is valid only when the local magnetic moment exists. Otherwise there is no way to discuss a specific spin dynamics separated from general dynamics of electron-hole excitations. In this section we derive the corresponding condition for the formation of the local magnetic moment in the system.

According to Landau phenomenology~\cite{LL_V}, a transition from a paramagnetic to a magnetically ordered state occurs due to a spontaneous symmetry breaking. 
The latter results in the change of the free energy ${\cal F}[m]$ from a paraboloid-like form with a minimum at $m=0$ to a Mexican-hat potential characterized by a continuous set of minima at $m\neq0$.
This change in the free energy can be seen in the sign change of the second variation of the free energy $\partial^{2}_{m}{\cal F}[m]\Big|_{m=0}$ with respect to the corresponding order parameter $m$ (see e.g. Ref.~\onlinecite{PhysRevB.102.224423}). 
As an example, let us consider a half-filled Hubbard model on a three-dimensional (3D) cubic lattice, where the spontaneous symmetry breaking is associated with the formation of the antiferromagnetic (AFM) ordering with the wave vector $\vec{Q}=\{\pi, \pi, \pi\}$.
The free energy of our problem is given by the  action derived above~\eqref{eq:S5} that is written in terms of the physical bosonic variables $\rho^{\varsigma}$ describing fluctuations of charge and spin densities.
Thus, the second variation of the free energy with respect to the AFM order parameter $\rho^{s}_{Q,\omega=0}$ results in the inverse of the AFM susceptibility $X^{s}_{Q,\omega=0}$~\cite{PhysRevLett.121.037204} that becomes zero at the transition point
\begin{align}
-\frac{\partial^2{\cal S}[\rho^{s}]}{\partial\rho^{s}_{Q,\omega=0}\partial\rho^{s}_{-Q,\omega=0}} = \left(X^{s}_{Q,\omega=0}\right)^{-1} = \left(\chi^{s}_{\omega=0}\right)^{-1} - {\cal I}^{ss}_{Q,\omega=0} = 0
\label{eq:AFM_criterion}
\end{align}

Above the AFM phase boundary fluctuations of magnetic moments are uncorrelated at large distances, which means that the moments on different lattice sites fluctuate independently on each other, assuming that the distance between sites is larger than the magnetic correlation length. 
It can be expected that the formation of the local magnetic moment can be captured in the same way as the formation of the AFM ordering but looking at the corresponding {\it local} free energy.
Importantly, this local free energy is different from the one of the local reference system~\eqref{eq:action_imp}.
Indeed, the impurity problem ${\cal S}_{\rm imp}$ describes local correlation effects of both, itinerant electrons and local magnetic moments.
In order to isolate the energy related to the magnetic moment only, one has to find a way to subtract the contribution of itinerant electrons from the local free energy of the reference system~\eqref{eq:action_imp}.
As we argue in Appendix~\ref{app:local}, this procedure can be done by excluding non-local terms from Eq.~\eqref{eq:S3} and integrating out fermionic variables $f^{(*)}$. This procedure is reminiscent of the mapping of the $s\text{-}d$ model on the Anderson impurity model for the $d$ electrons~\cite{doi:10.1080/00018738300101581, hewson_1993}.
Let is emphasize again that the discussion of local moments and their separate dynamics makes sense only at time scales much larger than typical electron times, such as the inverse of the hopping amplitude, or $1/U$. 
As the result we get the local problem written in terms of only physical bosonic variables $\rho$
\begin{align} 
{\cal S}_{\rm loc}
= 
&-\Tr\ln \left[g^{-1}_{\tau\tau'} \delta^{\phantom{*}}_{\sigma\sigma'} + \int^{\beta}_{0} d\tau'' \sum_{\varsigma} \sigma^{\varsigma}_{\sigma\sigma'} \Lambda^{\varsigma}_{\tau\tau'\tau''} \, \rho^{\varsigma}_{i\tau''} \right] \notag\\
&- \frac12 \iint_{0}^{\beta} d\tau\,d\tau' \sum_{\varsigma}\rho^{\varsigma}_{\tau} \left[\chi^{\varsigma}\right]^{-1}_{\tau\tau'} \rho^{\varsigma}_{\tau'} 
\label{eq:S10}
\end{align}
Note that the derivation of this local problem does not rely on the saddle point approximation for rotation angles, because no transformation of fermionic variables to a rotating frame~\eqref{eq:Rmatrix} has been performed in this case.

In analogy to the formation to the AFM state the formation of the local magnetic moment in the system can be seen in the sign change of the second variation of the local action~\eqref{eq:S10} with respect to the local magnetic moment
\begin{align}
-\frac{\partial^2{\cal S}_{\rm loc}[\rho^{s}]}{\partial\rho^{s}_{\tau}\,\partial\rho^{s}_{\tau'}} = \left[\chi^{s}\right]^{-1}_{\tau\tau'} - J^{\rm loc}_{\tau\tau'}
\label{eq:local_criterion}
\end{align}
Importantly, and contrary to the case of the true phase transitions for the infinite system, we keep times $\tau$ and $\tau'$ different since in the static limit local magnetic moment does not exist, it is screened by Kondo effect, or by intersite exchange-induced spin flips (in paramagnetic phase), or by both these factors. At the same time, as was already stressed, local magnetic moment exists at relatively long times in comparison with basic electron processes. In this sense, its existence means symmetry breaking at {\it intermediate} time scales.

\begin{figure}[t!]
\includegraphics[width=0.8\linewidth]{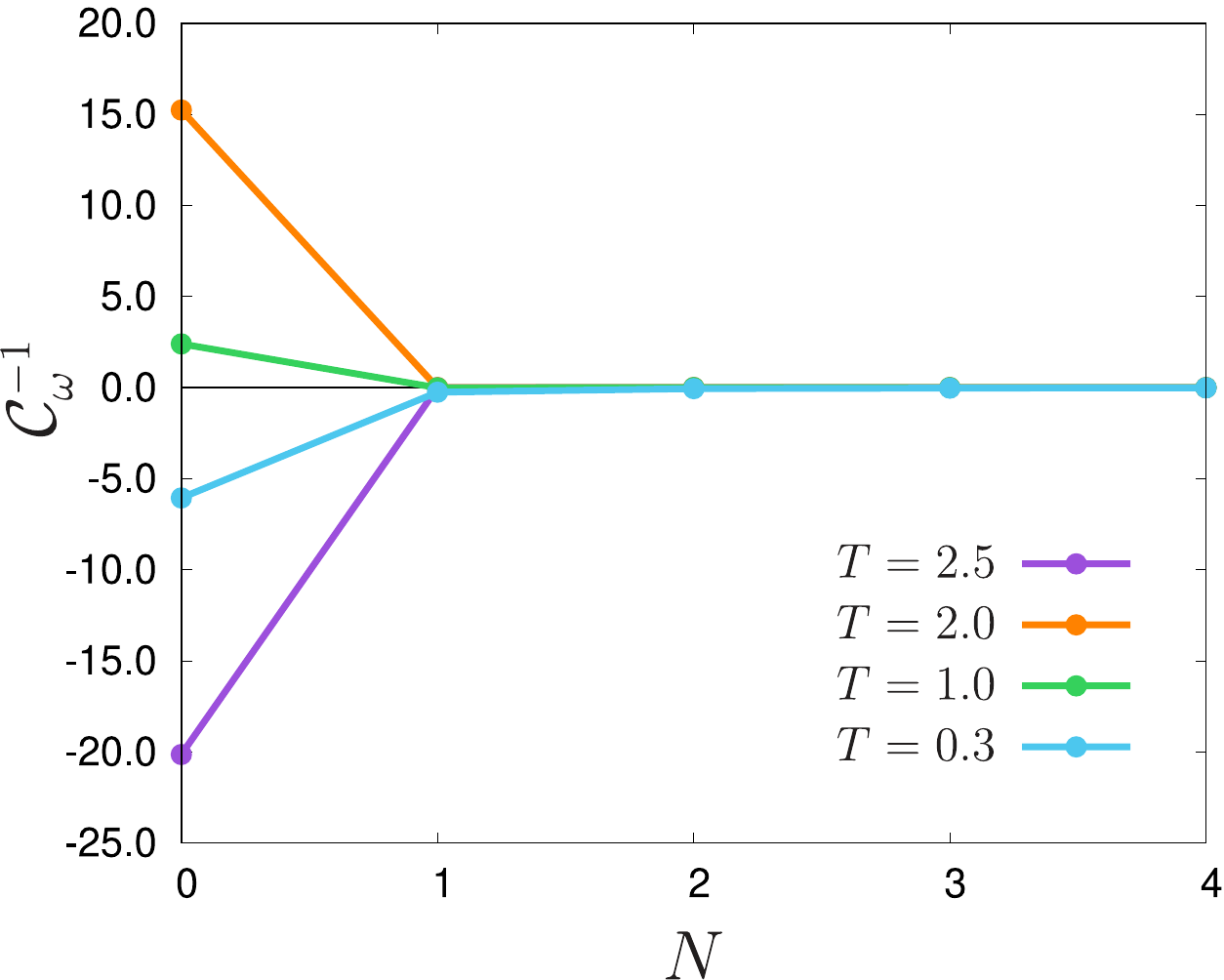}
\caption{\label{fig:criterion} ${\cal C}^{-1}_{\omega}$ as a function of the number $N$ of bosonic Matsubara frequency $\omega_{N}=2\pi{}N/\beta$. Results are obtained for $U=8$ in a broad range of temperatures specified in the legend. ${\cal C}_{\omega}$ has an opposite sign in regions characterized by zero and non-zero local magnetic moment and is nearly a delta-function in the frequency space.}
\end{figure}

The expression (\ref{eq:local_criterion}) corresponds to a ``slow'' exchange coupling of the local moment to itself at a different time point that can be obtained by subtracting the contribution of itinerant electrons given by a local analog of the exchange interaction~\eqref{eq:J}
\begin{align}
J^{\rm loc}_{\tau\tau'} = 
\int^{\beta}_{0} \{d\tau_i\} \sum_{\sigma} \Lambda^{\hspace{-0.05cm}*\,s}_{\tau\tau_1\tau_2} g^{\sigma}_{\tau_1\tau_3} g^{\sigma}_{\tau_4\tau_2} \Lambda^{\hspace{-0.05cm}s}_{\tau_3\tau_4\tau'}
\label{eq:J_loc}
\end{align}
from the total exchange interaction $\chi^{-1}_{\tau\tau'}$ of the local reference system~\eqref{eq:action_imp}. 
The moment when this self-exchange becomes diamagnetic clearly marks the instability of the truly paramagnetic phase without a developed local moment.
Remarkably, by direct numerical calculations we find that the self-interaction of the local moment~\eqref{eq:local_criterion} evaluated at equal times ${\tau=\tau'}$ never changes sign, at least for the considered model.
This fact suggests that the formation of the local magnetic moment is not a real physical transition and should rather be seen as a crossover.
In other words, although the electronic response is fast in time, electrons cannot react immediately to the polarization created by themselves.
In confirmation of that we find that at the critical temperature the second variation of the local free energy~\eqref{eq:local_criterion} changes sign at any times except ${\tau=\tau'}$.
In this sense the formation of the local moment drastically differs from the formation of the AFM ordering on a lattice that fulfills an equal-time criterion~\eqref{eq:AFM_criterion}. 
To illustrate this point, let us separate the static contribution in Eq.~\eqref{eq:local_criterion} that is contained in the inverse of the local susceptibility ${\chi^{s\,-1}_{\tau\tau'} = [\Pi^{s\,{\rm imp}}]^{-1}_{\tau\tau'} - \delta_{\tau\tau'}U^{s}}$.
In this expression $\Pi^{s\,{\rm imp}}_{\tau\tau'}$ is the polarization operator of the impurity problem~\eqref{eq:action_imp} and ${U^{s}=-U/2}$ is the bare interaction in the spin channel.
Then, the Eq.~\eqref{eq:local_criterion} becomes 
\begin{align}
-\partial^2_{\rho^{s}}{\cal S}_{\rm loc}[\rho^{s}] = {\cal C}_{\tau\tau'} - \delta_{\tau\tau'}U^{s}
\label{eq:local_criterion2}
\end{align}
where ${{\cal C}_{\tau\tau'} = [\Pi^{s\,{\rm imp}}]^{-1}_{\tau\tau'}} - J^{\rm loc}_{\tau\tau'}$ is an effective dynamical self-exchange coupling of the local magnetic moment.
The Fourier transform of ${{\cal C}^{-1}(\tau-\tau')}$ to Matsubara frequency $\omega$ space is shown in Fig.~\ref{fig:criterion}.
This quantity has a delta-functional behavior for all considered temperatures, which means that the second variation of the local action~\eqref{eq:local_criterion2} is nearly constant in time except a delta-function peak at ${\tau=\tau'}$ due to $\delta_{\tau\tau'}U^{s}$ term.
Therefore, the criterion for the local moment formation can be obtained by excluding this static contribution $\delta_{\tau\tau'}U^{s}$ from  Eq.~\eqref{eq:local_criterion2}, which results in the following condition
\begin{align}
{\cal C}_{\omega=0}=0
\label{eq:local_condition}
\end{align}
that is more convenient to write in the frequency space.
As we show in Appendix~\ref{app:Sigma}, the sign change of the self-exchange coupling ${\cal C}_{\omega}$ can be related to the sign change of the first variation of the local electronic self-energy with respect to the magnetization. 
The negative value of the variation indicates that the formation of the local magnetic moment is energetically favorable, because it minimizes the energy of electrons.
This condition can then serve as an approximate criterion for the formation of the local magnetic moment in the system.

\section{Application to 3D Hubbard model}
\label{sec:results}

\begin{figure}[t!]
\includegraphics[width=1.0\linewidth]{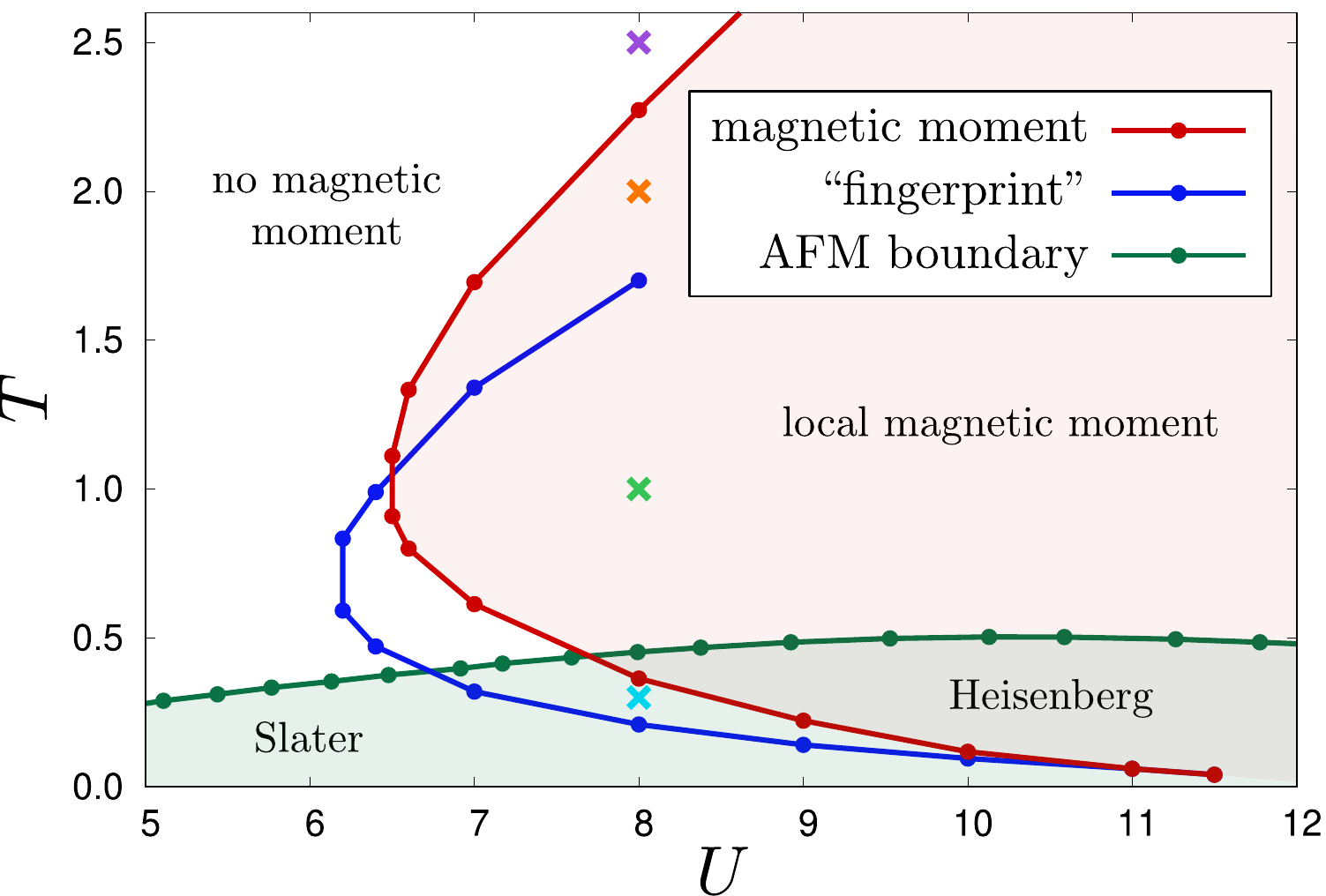}
\caption{\label{fig:phase} Phase diagram for the 3D Hubbard model as a function of temperature $T$ and local Coulomb interaction $U$. Red line corresponds to the criterion~\eqref{eq:local_condition} for the formation of the local magnetic moment. Blue line depicts the temperature at which the fingerprint of the local moment formation appears in the generalized local charge susceptibility (see Ref.~\cite{PhysRevLett.126.056403}). Green line is the AFM phase transition boundary obtained in Ref.~\cite{PhysRevB.92.144409} via DMFT susceptibility. Colored ``$\times$'' markers at $U=8$ highlight points for which the local free energy is shown in Fig.~\ref{fig:energy}.}
\end{figure}

In this section we explore the formation of the local magnetic moment in the context of the 3D Hubbard model.
For simplicity, we consider only the nearest-neighbor hopping amplitude $\varepsilon_{\av{ij}}=1$ which defines the energy unit of the system.
The SOC $\kappa_{ij}$ and the non-local interaction $V_{ij}$ are set to zero. 
Numerical results are obtained on the basis of the converged DMFT solution of the lattice problem~\eqref{eq:action_latt}.
In this case the local reference system is given by the impurity problem of DMFT~\eqref{eq:action_imp}.
Fig.~\ref{fig:phase} shows the phase diagram for the considered model in the temperature $(T)$ vs local Coulomb interaction $(U)$ space.
The red line corresponds to the condition~\eqref{eq:local_condition} that defines the temperature at which the local magnetic moment is formed in the system.
For illustrative purposes we also show the AFM phase boundary of DMFT (green line) that was obtained in Ref.~\cite{PhysRevB.92.144409} using the criterion on the inverse of the AFM susceptibility~\eqref{eq:AFM_criterion}.
Remarkably, we find that the local moment appears already at high temperatures well above the N\'eel phase transition to the AFM state.
This observation is consistent with the fact that in some cases a paramagnetic phase can be well described by a classical Heisenberg model that explicitly assumes the existence of the magnetic moment. 
At the same time, the obtained result suggests that the local magnetic moment exists only above a relatively large critical value of the local Coulomb interaction ${U^{*}\simeq6.5}$, which is comparable to the half of the bandwidth ${D/2=6}$.
More interestingly, the red line in Fig.~\ref{fig:phase} crosses the AFM phase boundary at even larger value of ${U\simeq7.7}$ and further splits the AFM phase into two parts.
The shaded red area to the right of the red line indicates the existence of the local magnetic moment in the system, which below the N\'eel temperature corresponds to Heisenberg antiferromagnetism (shaded green area).
Consequently, the AFM ordering to the left of the red line occurs without the presence of the magnetic moment, which can be associated with Slater mechanism of metal-insulator transition~\cite{PhysRev.82.538, PhysRevB.94.125144} due to long-range fluctuations of itinerant electrons. 
Remarkably, we find that the Slater (that is, purely itinerant) antiferromagnetism is not limited to a weakly-interacting regime of the Hubbard model and extends to large values of the local Coulomb interaction upon lowering the temperature. 

\begin{figure}[t!]
\includegraphics[width=1.0\linewidth]{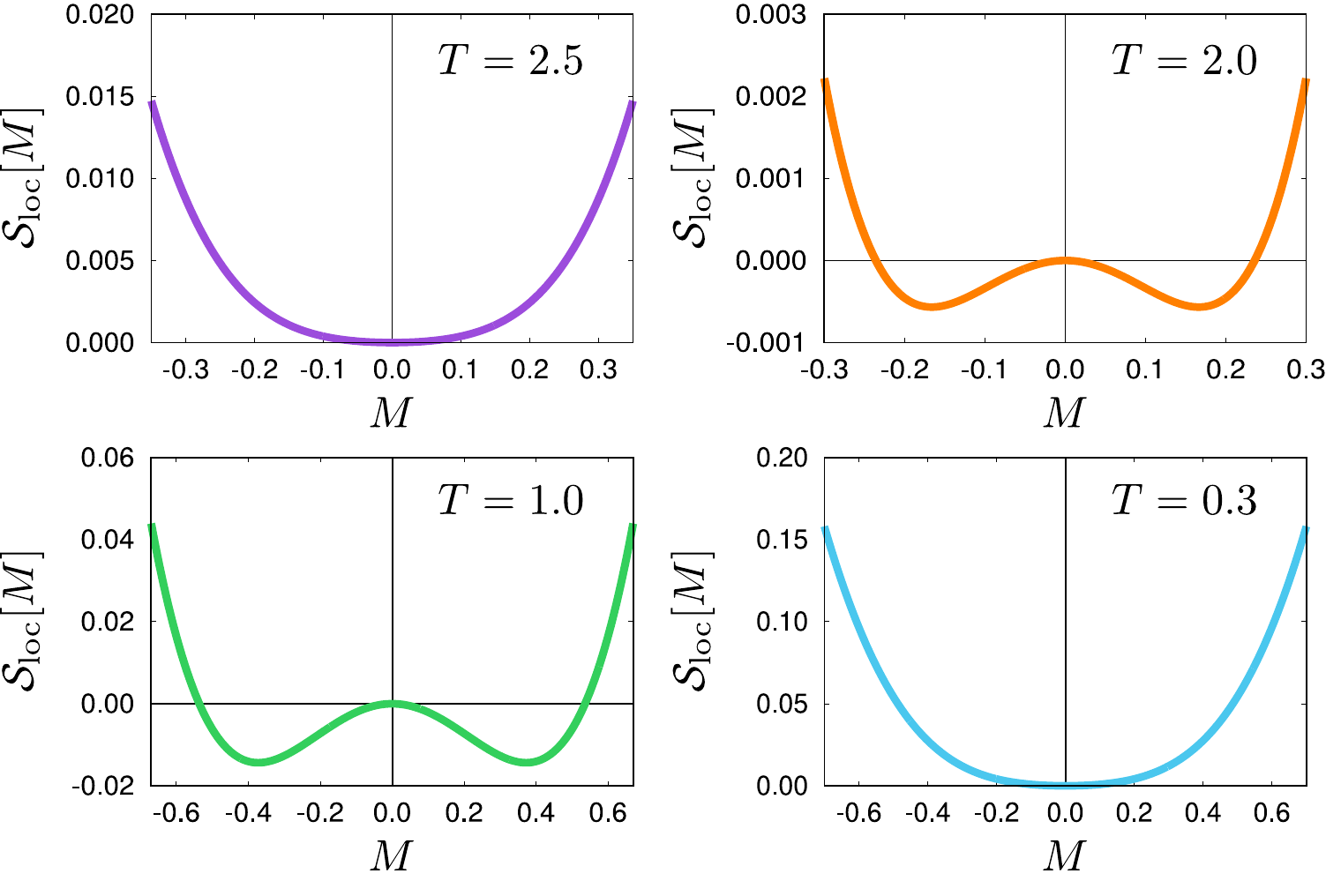}
\caption{\label{fig:energy} Local free energy of the system ${\cal S}_{\rm loc}[M]$ as a function of the local magnetic moment $M=\rho^{s}_{\omega=0}$. Results are obtained for $U=8$ at different temperatures $T=2.5$ (top left panel), $T=2.0$ (top right panel), $T=1.0$ (bottom left panel), and $T=0.3$ (bottom right panel). Color of each plot corresponds to the color of the ``$\times$'' marker in Fig.~\ref{fig:phase} that indicates the temperature at which the energy is calculated.}
\end{figure}

The average value of the local magnetic moment $\av{M}$ can be found from the local free energy of the system.
To this aim we exclude the static contribution from the inverse of the local susceptibility in Eq.~\eqref{eq:S10} according to discussions presented at the end of Section~\ref{sec:local_moment}, and plot the energy as a function of $M=\rho^{s}_{\omega=0}$.
Fig.~\ref{fig:energy} illustrates the corresponding result obtained for ${U=8}$ at different temperatures. 
We find that at a high temperature ${T=2.5}$ the local energy shown in top left panel has a parabolic form with a minimum at ${\av{M}=0}$.
After crossing the red line that depicts the formation of the local moment the energy takes the form of the double-well potential.
At ${T=2.0}$ two minima of the energy (top right panel) are located at $|{\av{M}|=0.17}$, which corresponds to the average value of the local magnetic moment. 
Upon lowering the temperature the value of the local magnetic moment grows and at ${T=1.0}$ becomes ${|\av{M}|=0.37}$, as shown in bottom left panel of Fig.~\ref{fig:energy}.
Finally, at ${T=0.3}$ (bottom right panel) after crossing the red line for the second time the minimum of the energy again shifts to  ${\av{M}=0}$, which corresponds to the Kondo screening of the local moment \cite{Hewson_book}. At small enough $U$ we rather deal with the regime of local spin fluctuations than with the Kondo effect; within Anderson model it corresponds to the regime of valence fluctuations~\cite{Hewson_book}. 
Free energy obtained for smaller (${U=6.4}$, ${U=6.8}$) and larger ($U=10$) values of the local Coulomb interaction close to a transition point can be found in Appendix~\ref{app:local}.

The average value of the local moment can also be compared to the magnetization $\av{M'}$ estimated from the equal-time local spin susceptibility (see e.g. Refs.~\onlinecite{PhysRevLett.104.197002, PhysRevB.86.064411, PhysRevLett.125.086402})
\begin{align}
3\chi^{s}_{\tau\tau} = \av{M'^2} \simeq \av{M'}\Big(\av{M'}+2 \, \Big)
\end{align}
This expression gives nearly constant value $\av{M'}=0.85\pm0.03$ for the all considered temperatures.
Therefore, we find that the magnetization calculated from the susceptibility strongly overestimates the average value of the local magnetic moment $\av{M}$ and, moreover, remains very large in the regime where the system possesses no local magnetic moment at all.
This result is a direct consequence of the fact that the spin susceptibility of the local reference system~\eqref{eq:action_imp} cannot distinguish the fluctuations of the local magnetic moment from the spin fluctuations of the itinerant electrons that also contribute to the susceptibility, especially in the paramagnetic regime~\cite{moriya2012spin}.
In particular, this explains the large value of $\av{M'}$ obtained at $T=0.3$ that lies in the Kondo (or valence fluctuation) regime, where the local magnetic moment is screened by the electrons. 

Importantly, the Kondo screening of the local magnetic moment can influence different physical quantities not necessarily related to magnetism.
Recently, it has been argued that the appearance of the ``onion-like'' fermionic Matsubara frequency structure of the generalized local charge susceptibility ${\chi^{\rm c}(\nu,\nu',\omega)}$ can be seen as a fingerprint of the Kondo screening of the local magnetic moment~\cite{PhysRevLett.126.056403}.
In particular, the authors of Ref.~\onlinecite{PhysRevLett.126.056403} showed that in the low temperature and large $U$ regime the criterion ${\chi^{\rm c}(\nu_0,\nu_0,\omega_0)=\chi^{\rm c}(\nu_0,-\nu_0,\omega_0)}$ for this change in the fermionic frequency structure, where ${\nu_0=\pi/\beta}$ and ${\omega_0=0}$ are respectively the zeroth fermionic and bosonic Matsubara frequencies, provides a good estimate for the Kondo temperature.
For comparison we plot the ``fingerprint'' temperature (blue line in Fig.~\ref{fig:phase}) that was deduced from the criterion on the local generalized charge susceptibility of the DMFT impurity problem~\eqref{eq:action_imp}.
We find that in the large interaction regime ${U\geq11}$ the lower branches of both red and blue curves lie on top of each other and thus reproduce the Kondo temperature obtained in Ref.~\onlinecite{PhysRevLett.126.056403}.
However, upon decreasing the interaction the blue line rapidly shifts downwards to lower temperatures from the red one. This trend is preserved and significantly increased upon going to the high-temperature branches. 
Therefore, we observe that the phenomenological criterion proposed in Ref.~\onlinecite{PhysRevLett.126.056403} seems to correlate with the formation of the local moment but can be  directly attributed to it only in the well-developed Kondo regime. 

\section{Conclusions}

To conclude, in this work we have shown that the extended Hubbard model that describes the behavior of electrons on a lattice can be turned into an effective bosonic action describing spin dynamics of itinerant-electron systems.
We have proved that the special set of transformations allows one to write the resulting action in terms of physical variables that can be associated with fluctuations of the local magnetic moment. 
In particular, the use of these transformations makes possible a careful separation of electronic and spin degrees of freedom.
We have shown that the derived effective spin problem takes into account all possible spin exchange interactions including the isotropic, the antisymmetric anisotropic, and the symmetric anisotropic interactions.
Importantly, the obtained two-spin interactions can be reduced to a well-known expression \cite{LKG84, LKAG87, KL2000} that is intensively used for realistic calculations of magnetic materials.
Another important step in these transformations is a precise separation of the rotational dynamics of the local magnetic moment from the Higgs fluctuations of its modulus.
This allowed us to obtain a correct Berry phase term, which is written in terms of rotation angles and of the scalar field that describes the behavior of the modulus of the magnetic moment.
As a consequence, the equation of motion for the introduced spin problem takes the Landau-Lifshitz-Gilbert form, which, in addition to the uniform spin precession, accounts for the Gilbert damping.

The derived equation of motion assumes the presence of the local magnetic moment in the system.
To capture the formation of the magnetic moment we have investigated the local free energy of the system.
We have found that, according to Landau phenomenology, the energy as a function of the local magnetic moment transforms from the parabolic-like to a Mexican-hat potential form.
The transition point between two cases can be captured by a sign change of the second variation of the local free energy, which allowed us to introduce a criterion for the formation of the local magnetic moment.
This criterion can be interpreted as the appearance of a diamagnetic exchange coupling of the local magnetic to itself.
According to the derived criterion this self-exchange coupling can be obtained by excluding the itinerant contribution from the total local exchange.
This result explains the fact that the value of the local magnetic moment cannot be estimated from the local magnetic susceptibility, which also contains the contribution of itinerant electrons.
Instead, we have argued that the value of the local magnetic moment can be found from the minima of the local free energy.
As the result, we have observed that the local moment appears in the system well above the transition temperature to the ordered magnetic state and exists only above a critical value of the Coulomb interaction.

\begin{acknowledgments}
The authors are thankful to Alexey Rubtsov, Leonid Pourovskii, and Alessandro Toschi for useful discussions and comments. 
The work of E.A.S. was supported by the European Union’s Horizon 2020 Research and Innovation programme under the Marie Sk\l{}odowska Curie grant agreement No.~839551 - $\text{2DMAGICS}$. The work of M.I.K. and A.I.L. was supported by European Research Council via Synergy Grant 854843 - FASTCORR. S.B., V.H., and A.I.L. acknowledge the support by the Cluster of Excellence ``Advanced Imaging of Matter'' of the Deutsche Forschungsgemeinschaft (DFG) - EXC 2056 - Project No.~ID390715994. E.A.S., V.H., and A.I.L. also acknowledge the support by North-German Supercomputing Alliance (HLRN) under the Project No.~hhp00042.\\
\end{acknowledgments}

\onecolumngrid
\appendix

\section{Effective spin problem in the multiorbital case}
\label{app:HS}

In this section we explicitly derive a spin problem for a multi-orbital ($ll'$) electronic system. 
Those who are not interested in technical details of this derivation, can find a summary of performed transformations at the end of this section.
We start with the lattice action of the extended Hubbard model written in the coordinate ($i$) and imaginary time ($\tau$) space
\begin{align}
{\cal S} = \int_{0}^{\beta} d\tau \, \Bigg\{  -\hspace{-0.2cm}\sum_{ij,\sigma\sigma',ll'} \hspace{-0.1cm} c^{*}_{i\tau\sigma{}l} \left[\delta_{ij}\delta_{\sigma\sigma'}\delta_{ll'}(-\partial_{\tau}+\mu)-\varepsilon^{\sigma\sigma'}_{ij,ll'}\right]
c_{j\tau\sigma'l'}^{\phantom{*}}
+ \frac12 \sum_{i,\sigma\sigma',\{l\}} \hspace{-0.1cm} U^{\phantom{*}}_{l_1 l_2 l_3 l_4} c^{*}_{i\tau\sigma{}l_1} c^{\phantom{*}}_{i\tau\sigma{}l_2} c^{*}_{i\tau\sigma'{}l_4} c^{\phantom{*}}_{i\tau\sigma'{}l_3}
+\frac12 \sum_{ij,\varsigma,\{l\}}  \rho^{\varsigma}_{i\tau{}l_1l_2} V^{\varsigma,\,ij}_{l_1l_2l_3l_4} \rho^{\varsigma}_{j\tau{}l_4l_3} \Bigg\} \label{eq:action_latt_app}
\end{align}
In this expression we introduced composite variables ${\rho^{\varsigma}_{i\tau{}ll'} = n^{\varsigma}_{i\tau{}ll'} - \av{n^{\varsigma}_{ll'}}}_{\rm imp}$ that describe fluctuations of the charge (${\varsigma=c}$) and spin (${\varsigma=s=\{x,y,z\}}$) densities ${n^{\varsigma}_{i\tau{ll'}} = \sum_{\sigma\sigma'} c^{*}_{i\tau\sigma{}l} \, \sigma^{\varsigma}_{\sigma\sigma'} c^{\phantom{*}}_{i\tau\sigma'l'}}$ around their average values. 
The latter are provided by an effective site-independent impurity problem
\begin{align}
{\cal S}_{\rm imp} = &- \iint_{0}^{\beta} d\tau\,d\tau'
\sum_{\sigma,ll'} c^{*}_{\tau\sigma{}l} \left[\delta_{\tau\tau'}\delta_{ll'}(-\partial_{\tau}+\mu) - \Delta^{ll'}_{\tau\tau'}\right] c^{\phantom{*}}_{\tau'\sigma{}l'}
+\frac12 \int_{0}^{\beta} d\tau \sum_{\sigma\sigma',\{l\}} U^{\phantom{*}}_{l_1 l_2 l_3 l_4} c^{*}_{\tau\sigma{}l_1} c^{\phantom{*}}_{\tau\sigma{}l_2} c^{*}_{\tau\sigma'{}l_4} c^{\phantom{*}}_{\tau\sigma'{}l_3}
\label{eq:action_imp_app}
\end{align}
The impurity problem is considered in a non-polarized form so that $\av{n^{s}_{ll'}}_{\rm imp}=0$.
Consequently, the corresponding retarded fermionic 
hybridization function is spin independent and reads ${\Delta^{ll'}_{\tau\tau'} = \Delta^{ll'}(\tau-\tau')}$. 
The remainder part of the action ${{\cal S}_{\rm rem} = {\cal S} - \sum_{i} {\cal S}_{\rm imp}}$ is following
\begin{align}
{\cal S}_{\rm rem} = \iint_{0}^{\beta} d\tau\, d\tau'
\sum_{ij,\sigma\sigma',ll'} c^{*}_{i\tau\sigma{}l}\,
\tilde{\varepsilon}^{\tau\tau'll'}_{ij\sigma\sigma'}\,
c^{\phantom{*}}_{j\tau'\sigma'l'}
+\frac12\int_{0}^{\beta} d\tau \sum_{ij,\varsigma,\{l\}} \rho^{\varsigma}_{i\tau{}l_1l_2} V^{\varsigma,\,ij}_{l_1l_2l_3l_4} \, \rho^{\varsigma}_{j\tau{}l_4l_3}
\label{eq:action_rem_app}
\end{align}
where $\tilde{\varepsilon}^{\tau\tau'll'}_{ij\sigma\sigma'} =  \varepsilon^{\sigma\sigma'}_{ij,ll'}\delta^{\phantom{*}}_{\tau\tau'} - \Delta_{\tau\tau'}^{ll'}\delta^{\phantom{*}}_{ij}\delta^{\phantom{*}}_{\sigma\sigma'}$.
In order to integrate out the local reference system~\eqref{eq:action_imp_app} that is solved numerically exactly we perform Hubbard-Stratonovich transformations over the remainder part of the lattice action~\eqref{eq:action_rem_app} as
\begin{align}
&\exp\left\{ - \iint_{0}^{\beta} d\tau \, d\tau' \sum_{ij,\sigma\sigma',ll'} c^{*}_{i\tau\sigma{}l}\,
\tilde{\varepsilon}^{\tau\tau'll'}_{ij\sigma\sigma'}\,
c^{\phantom{*}}_{j\tau'\sigma'l'}\right\} 
= \notag\\
&{\cal D}_{f} \int D[f^{*},f] \exp\left\{ \int_{0}^{\beta} \{d\tau_{i}\} \sum_{ij,\sigma\sigma',\{l\}} f^{*}_{i\tau_1\sigma{}l_1} \left[g^{-1}\right]^{l_1l_2}_{\tau_1\tau_2}
\left[\tilde{\varepsilon}^{-1}\right]^{\tau_2\tau_3,l_2l_3}_{ij,\sigma\sigma'}\left[g^{-1}\right]^{l_3l_4}_{\tau_3\tau_4}
f^{\phantom{*}}_{j\tau_4\sigma'l_4} 
- \iint_{0}^{\beta} d\tau\, d\tau' \sum_{i,\sigma,ll'}\left( c^{*}_{i\tau\sigma{}l}\left[g^{-1}\right]^{ll'}_{\tau\tau'}f^{\phantom{*}}_{i\tau'\sigma{}l'} + h.c.
\right)\right\} 
\label{eq:HSf1_app}\\
&\exp\left\{-\frac12\int_{0}^{\beta} d\tau \sum_{ij,\varsigma,\{l\}} \rho^{\varsigma}_{i\tau{}l_1l_2} V^{\varsigma,\,ij}_{l_1l_2l_3l_4} \,
\rho^{\varsigma}_{j\tau{}l_4l_3}\right\} 
={\cal D}_{\phi}
\int D[\phi^{\varsigma}] \exp\left\{ \int_{0}^{\beta} d\tau \, \left( \frac12\sum_{ij,\varsigma,\{l\}}\phi^{\varsigma}_{i\tau{}l_1l_2} \left[\left(V^{\varsigma}\right)^{-1}\right]^{ij}_{l_1l_2,l_3l_4} \phi^{\varsigma}_{j\tau{}l_4l_3} -
\sum_{i,\varsigma,ll'}\phi^{\varsigma}_{i\tau{}ll'} \, \rho^{\varsigma}_{i\tau{}l'l}\right)\right\}
\label{eq:HSb1_app}
\end{align}
where determinants ${\cal D}_{f} = {\rm det}\left[g(\Delta-\varepsilon)g\right]$ and ${\cal D}^{-1}_{\phi} = -\sqrt{{\rm det}V}$ can be neglected when calculating expectation values. 
After that the lattice action becomes
\begin{align} 
{\cal S}
= 
&-\iint_{0}^{\beta} \{d\tau_{i}\} \sum_{ij,\sigma\sigma',\{l\}} f^{*}_{i\tau_1\sigma{}l_1} \left[g^{-1}\right]^{l_1l_2}_{\tau_1\tau_2}
\left[\tilde{\varepsilon}^{-1}\right]^{\tau_2\tau_3,l_2l_3}_{ij,\sigma\sigma'}\left[g^{-1}\right]^{l_3l_4}_{\tau_3\tau_4}
f^{\phantom{*}}_{j\tau_4\sigma'l_4} 
+\iint_{0}^{\beta} d\tau\, d\tau' \sum_{i,\sigma,ll'}\left( c^{*}_{i\tau\sigma{}l}\left[g^{-1}\right]^{ll'}_{\tau\tau'}f^{\phantom{*}}_{i\tau'\sigma{}l'} + f^{*}_{i\tau\sigma{}l}\left[g^{-1}\right]^{ll'}_{\tau\tau'}c^{\phantom{*}}_{i\tau'\sigma{}l'}
\right)
\notag\\
&-\frac12\int_{0}^{\beta} d\tau \sum_{ij,\varsigma,\{l\}}\phi^{\varsigma}_{i\tau{}l_1l_2} \left[\left(V^{\varsigma}\right)^{-1}\right]^{ij}_{l_1l_2,l_3l_4} \phi^{\varsigma}_{j\tau{}l_4l_3}
+\int_{0}^{\beta} d\tau \sum_{i,\varsigma,ll'} 
\left(\phi^{\varsigma}_{i\tau{}ll'} + j^{\,\varsigma}_{i\tau{}ll'}\right) \rho^{\varsigma}_{i\tau{}l'l}
+ \sum_{i}{\cal S}_{\rm imp}
\label{eq:S_HS_app}
\end{align}
In the last equation we have introduced source field
$j^{\,\varsigma}$ for the composite $\rho^{\varsigma}$ variable.
Now we perform a transformation of fermionic variables to a rotating frame as discussed in the main text. 
This transformation can be written in terms of the unitary orbital-independent rotation matrix
\begin{align}
R_{i\tau} = 
\begin{pmatrix}
\cos(\theta_{i\tau}/2) & - e^{-i\varphi_{i\tau}}\sin(\theta_{i\tau}/2) \\
e^{i\varphi_{i\tau}}\sin(\theta_{i\tau}/2) & \cos(\theta_{i\tau}/2)
\end{pmatrix}
\label{eq:Rmatr_app}
\end{align}
where polar angles $\theta_{i\tau}$ and $\varphi_{i\tau}$ will be defined later.
Fermionic variables are transformed as follows 
${c_{i\tau{}l} \to R_{i\tau} c_{i\tau{}l}}$,
where ${c_{i\tau{}l} = (c_{i\tau{}l\uparrow}, c_{i\tau{}l\downarrow})^{T}}$. The action becomes
\begin{align} 
{\cal S} 
= 
&-\iint_{0}^{\beta} \{d\tau_{i}\} \sum_{ij,\sigma\sigma',\{l\}} f^{*}_{i\tau_1\sigma{}l_1} \left[g^{-1}\right]^{l_1l_2}_{\tau_1\tau_2}
\left[\tilde{\varepsilon}^{-1}\right]^{\tau_2\tau_3,l_2l_3}_{ij,\sigma\sigma'}\left[g^{-1}\right]^{l_3l_4}_{\tau_3\tau_4}
f^{\phantom{*}}_{j\tau_4\sigma'l_4}
+ \iint_{0}^{\beta} d\tau \, d\tau' \Tr_{\sigma}\sum_{i,ll'} \Bigg\{ c^{*}_{i\tau{}l} R^{\dagger}_{i\tau} \left[g^{-1}\right]^{ll'}_{\tau\tau'} f^{\phantom{*}}_{i\tau'l'} + f^{*}_{i\tau{}l} \left[g^{-1}\right]^{ll'}_{\tau\tau'} R^{\phantom{\dagger}}_{i\tau'} c^{\phantom{*}}_{i\tau'l'}\Bigg\}
\notag\\
&-\frac12\int_{0}^{\beta} d\tau \sum_{ij,\varsigma,\{l\}} 
\phi^{\varsigma}_{i\tau{}l_1l_2}
\left[\left(V^{\varsigma}\right)^{-1}\right]^{ij}_{l_1l_2,l_3l_4} 
\phi^{\varsigma}_{j\tau{}l_4l_3}
+\int_{0}^{\beta} d\tau \sum_{i,ll'} 
\left(\phi^{c}_{i\tau{}ll'} + j^{c}_{i\tau{}ll'}\right) \rho^{c}_{i\tau{}l'l}
+ \int_{0}^{\beta} d\tau \,\Tr_{\sigma}\sum_{i,s,ll'} \left(\phi^{s}_{i\tau{}ll'} + j^{s}_{i\tau{}ll'}\right) c^{*}_{i\tau{}l'} R^{\dagger}_{i\tau} \sigma^{s} R^{\phantom{*}}_{i\tau} c^{\phantom{*}}_{i\tau{}l} \notag\\
&+ \int_{0}^{\beta} d\tau \,\Tr_{\sigma}\sum_{i,l} c^{*}_{i\tau{}l} R^{\dagger}_{i\tau} \dot{R}^{\phantom{*}}_{i\tau} c^{\phantom{*}}_{i\tau{}l} 
+ \sum_{i}{\cal S}_{\rm imp}
\label{eq:action_pp}
\end{align}
The term that contains the derivative of the rotation matrix can be rewritten as an effective gauge field  
${R^{\dagger}_{i\tau} \dot{R}^{\phantom{*}}_{i\tau} = \sum_{s} {\cal A}^{s}_{i\tau}\sigma^{s}}$.
We also rewrite the coupling of an effective magnetic field to electronic degrees of freedom as
${R^{\dagger}_{i\tau} \sigma^{\varsigma} R^{\phantom{*}}_{i\tau} 
= \sum_{\varsigma'} {\cal U}^{\varsigma\varsigma'}_{i\tau}\sigma^{\varsigma'}}$,
where ${[{\cal U}_{i\tau}^{-1}]^{ss'}=[{\cal U}^{\rm T}_{i\tau}]^{ss'}}$, ${{\cal U}^{cs}_{i\tau}=0}$, and ${{\cal U}^{cc}_{i\tau}=1}$.
Using these notations the action can be rewritten as
\begin{align} 
{\cal S} 
= 
&-\iint_{0}^{\beta} \{d\tau_{i}\} \sum_{ij,\sigma\sigma',\{l\}} f^{*}_{i\tau_1\sigma{}l_1} \left[g^{-1}\right]^{l_1l_2}_{\tau_1\tau_2}
\left[\tilde{\varepsilon}^{-1}\right]^{\tau_2\tau_3,l_2l_3}_{ij,\sigma\sigma'}\left[g^{-1}\right]^{l_3l_4}_{\tau_3\tau_4}
f^{\phantom{*}}_{j\tau_4\sigma'l_4}
+ \iint_{0}^{\beta} d\tau \, d\tau' \Tr_{\sigma}\sum_{i,ll'} \Bigg\{ c^{*}_{i\tau{}l} R^{\dagger}_{i\tau} \left[g^{-1}\right]^{ll'}_{\tau\tau'} f^{\phantom{*}}_{i\tau'l'} + f^{*}_{i\tau{}l} \left[g^{-1}\right]^{ll'}_{\tau\tau'} R^{\phantom{\dagger}}_{i\tau'} c^{\phantom{*}}_{i\tau'l'}\Bigg\} \notag\\
&-\frac12\int_{0}^{\beta} d\tau \sum_{ij,\varsigma,\{l\}} 
\phi^{\varsigma}_{i\tau{}l_1l_2}
\left[\left(V^{\varsigma}\right)^{-1}\right]^{ij}_{l_1l_2,l_3l_4} 
\phi^{\varsigma}_{j\tau{}l_4l_3}
+ \int_{0}^{\beta} d\tau \sum_{i,\varsigma\varsigma',ll'}\left(\phi^{\varsigma}_{i\tau{}ll'} + j^{\,\varsigma}_{i\tau{}ll'} + \sum_{\varsigma''} {\cal A}^{\varsigma''}_{i\tau}[{\cal U}_{i\tau}^{-1}]^{\varsigma''\varsigma}\delta^{\phantom{\dagger}}_{ll'}\right) {\cal U}^{\varsigma\varsigma'}_{i\tau}\rho^{\varsigma'}_{i\tau{}l'l} 
+ \sum_{i}{\cal S}_{\rm imp}
\label{eq:action_pp2}
\end{align}
Now we make a shift of bosonic variables
${\phi^{\varsigma}_{i\tau{}ll'} \to  \hat{\phi}^{\varsigma}_{i\tau{}ll'} = \phi^{\varsigma}_{i\tau{}ll'} - j^{\,\varsigma}_{i\tau{}ll'} - \sum_{\varsigma''}{\cal A}^{\varsigma''}_{i\tau}[{\cal U}_{i\tau}^{-1}]^{\varsigma''\varsigma}\delta^{\phantom{\dagger}}_{ll'}}$
that leads to
\begin{align} 
{\cal S} 
= 
&-\iint_{0}^{\beta} \{d\tau_{i}\} \sum_{ij,\sigma\sigma',\{l\}} f^{*}_{i\tau_1\sigma{}l_1} \left[g^{-1}\right]^{l_1l_2}_{\tau_1\tau_2}
\left[\tilde{\varepsilon}^{-1}\right]^{\tau_2\tau_3,l_2l_3}_{ij,\sigma\sigma'}\left[g^{-1}\right]^{l_3l_4}_{\tau_3\tau_4}
f^{\phantom{*}}_{j\tau_4\sigma'l_4}
+ \iint_{0}^{\beta} d\tau \, d\tau' \Tr_{\sigma}\sum_{i,ll'} \Bigg\{ c^{*}_{i\tau{}l} R^{\dagger}_{i\tau} \left[g^{-1}\right]^{ll'}_{\tau\tau'} f^{\phantom{*}}_{i\tau'l'} + f^{*}_{i\tau{}l} \left[g^{-1}\right]^{ll'}_{\tau\tau'} R^{\phantom{\dagger}}_{i\tau'} c^{\phantom{*}}_{i\tau'l'}\Bigg\} \notag\\
&-\frac12\int_{0}^{\beta} d\tau \sum_{ij,\varsigma,\{l\}} \hat{\phi}^{\,\varsigma}_{i\tau{}l_1l_2} \left[\left(V^{\varsigma}\right)^{-1}\right]^{ij}_{l_1l_2,l_3l_4} \hat{\phi}^{\,\varsigma}_{j\tau{}l_4l_3}
+ \int_{0}^{\beta} d\tau \sum_{i,\varsigma\varsigma',ll'} \phi^{\varsigma}_{i\tau{}ll'} {\cal U}^{\varsigma\varsigma'}_{i\tau}\rho^{\varsigma'}_{i\tau{}l'l}
+ \sum_{i}{\cal S}_{\rm imp}
\label{eq:action_pp4}
\end{align}
After shifting bosonic variables the gauge ${\cal A}$ and the source $j$ fields are no longer coupled to original fermionic $c^{(*)}$ degrees of freedom and thus will not be affected by the integration of original fermionic degrees of freedom.
The latter is performed as
\begin{align}
&\int D[c^{*},c]\,\exp\left\{ -\sum_{i} {\cal S}_{\rm imp} 
-\iint_{0}^{\beta} d\tau \, d\tau' \Tr_{\sigma}\sum_{i,ll'} \Bigg\{ c^{*}_{i\tau{}l} R^{\dagger}_{i\tau} \left[g^{-1}\right]^{ll'}_{\tau\tau'} f^{\phantom{*}}_{i\tau'l'} + f^{*}_{i\tau{}l} \left[g^{-1}\right]^{ll'}_{\tau\tau'} R^{\phantom{\dagger}}_{i\tau'} c^{\phantom{*}}_{i\tau'l'}\Bigg\}
- \int_{0}^{\beta} d\tau \sum_{i,\varsigma\varsigma',ll'} \phi^{\varsigma}_{i\tau{}ll'} {\cal U}^{\varsigma\varsigma'}_{i\tau}\rho^{\varsigma'}_{i\tau{}l'l} \right\} 
= \notag\\
&{\cal Z}_{\rm imp} \times \exp\left\{
-\int_{0}^{\beta} \{d\tau_i\} \sum_{i,\{l\}} \left(
\Tr_{\sigma} 
f^{*}_{i\tau_1l_1} \left[g^{-1}\right]^{l_1l_2}_{\tau_1\tau_2} R^{\phantom{\dagger}}_{i\tau_2} g^{l_2l_3}_{\tau_2\tau_3} R^{\dagger}_{i\tau_3} \left[g^{-1}\right]^{l_3l_4}_{\tau_3\tau_4} f^{\phantom{*}}_{i\tau_4l_4} 
+ \frac12 \sum_{\{\varsigma\}} \phi^{\varsigma_1}_{i\tau_1l_1l_2} {\cal U}^{\varsigma_1\varsigma_2}_{i\tau_1} \chi^{\varsigma_2,\,\tau_1\tau_2}_{l_1l_2l_3l_4} [{\cal U}_{i\tau_2}^{-1}]^{\varsigma_2\varsigma_3} \phi^{\varsigma_3}_{i\tau_2l_4l_3} \right)
-\tilde{\cal F}[f,\phi] \right\}
\label{eq:Integrating_S_imp_app}
\end{align}
where ${\cal Z}_{\rm imp}$, $g_{\tau\tau'}^{ll'} = -\av{c^{\phantom{*}}_{i\tau{}l} \, c^{*}_{i\tau'l'}}$, and
$\chi^{\varsigma\,\tau\tau'}_{l_1 l_2 l_3 l_4} = - \langle \rho^{\varsigma}_{i\tau l_2 l_1} \, \rho^{\varsigma}_{i\tau' l_3 l_4} \rangle$ are the partition function, the Green's function, and the susceptibility of the local impurity problem~\eqref{eq:action_imp_app}, respectively.
The interaction $\tilde{\cal F}[f,\phi]$ that appears as the result of the integration of $c^{(*)}$ variables contains all possible exact fermion-fermion, fermion-boson, and boson-boson vertex functions of the local reference system~\eqref{eq:action_imp_app}. 
However, in actual calculations it is usually truncated at the two-particle level, which is a standard approximation that is widely used in the dual fermion~\cite{PhysRevB.77.033101, PhysRevB.79.045133, PhysRevLett.102.206401}, the dual boson~\cite{Rubtsov20121320, PhysRevB.90.235135, PhysRevB.93.045107, PhysRevB.94.205110, PhysRevB.100.165128}, and the dual TRILEX~\cite{PhysRevB.100.205115, PhysRevB.103.245123, 2020arXiv201003433S} methods including their diagrammatic Monte Carlo realizations~\cite{PhysRevB.94.035102, PhysRevB.96.035152, PhysRevB.102.195109}. Thus, the truncated interaction contains only the exact local three-point $\Lambda^{\varsigma,\,\tau_1\tau_2\tau_3}_{l_1l_2;l_3l_4}$ and four-point $\Gamma^{\tau_1\tau_2\tau_3\tau_4}_{l_1l_2l_3l_4}$ vertex functions of the reference system
\begin{align}
\tilde{\cal F}[f,\phi]
\simeq
\int_{0}^{\beta} \{d\tau_{i}\} \Tr_{\sigma} \sum_{i,\{l\}} \Bigg[
&\sum_{\varsigma\varsigma'}f^{*}_{i\tau_1l_1}
R^{\phantom{\dagger}}_{i\tau_1}\sigma^{\varsigma}R^{\dagger}_{i\tau_2}
f^{\phantom{*}}_{i\tau_2l_2} \Lambda^{\varsigma,\,\tau_1\tau_2\tau_3}_{l_1l_2;l_3l_4} \chi^{\varsigma,\,\tau_3\tau_4}_{l_3l_4l_5l_6} [{\cal U}_{i\tau_4}^{-1}]^{\varsigma\varsigma'} \phi^{\varsigma'}_{i\tau_4l_6l_5} 
+ \frac14 \Gamma^{\tau_1\tau_2\tau_3\tau_4}_{l_1l_2l_3l_4} f^{*}_{i\tau_1l_1}
R^{\phantom{\dagger}}_{i\tau_1} R^{\dagger}_{i\tau_2}
f^{\phantom{*}}_{i\tau_2l_2}f^{*}_{i\tau_4l_4}
R^{\phantom{\dagger}}_{i\tau_4} R^{\dagger}_{i\tau_3}
f^{\phantom{*}}_{i\tau_3l_3}
\Bigg]
\label{eq:lowestint}
\end{align}
where in the paramagnetic regime $\Lambda^{\hspace{-0.05cm}x}=\Lambda^{\hspace{-0.05cm}y}=\Lambda^{\hspace{-0.05cm}z}$, and
\begin{align}
\Lambda_{l_1l_2;l_3l_4}^{\hspace{-0.05cm}c/z,\,\tau_1\tau_2\tau_3} 
&= \int^{\beta}_{0} \{d\tau'\} \sum_{\{l'\}} \av{c^{\phantom{*}}_{i\tau'_1\uparrow{}l'_1} c^{*}_{i\tau'_2\uparrow{}l'_2} \, \rho^{c/z}_{i\tau'_3l'_3l'_4}} \left[g^{-1}\right]^{l'_1l_1}_{\tau'_1\tau_1} \left[g^{-1}\right]^{l'_2l_2}_{\tau'_2\tau_2} \left[\left(\chi^{c/z}\right)^{-1}\right]^{\tau'_3\tau_3}_{l'_3l'_4,l_3l_4} 
\label{eq:Lambda_app}\\
\Gamma^{\tau_1\tau_2\tau_3\tau_4}_{l_1l_2l_3l_4, \sigma_1\sigma_2\sigma_3\sigma_4} 
&= \int^{\beta}_{0} \{d\tau'\} \sum_{\{l'\}} \av{c^{\phantom{*}}_{i\tau'_1\sigma_1l'_1} c^{*}_{i\tau'_2\sigma_2l'_2} c^{*}_{i\tau'_3\sigma_3l'_3} c^{\phantom{*}}_{i\tau'_4\sigma_4l'_4}}_{\rm c} 
\left[g^{-1}\right]^{l'_1l_1}_{\tau'_1\tau_1} \left[g^{-1}\right]^{l'_2l_2}_{\tau'_2\tau_2} \left[g^{-1}\right]^{l'_3l_3}_{\tau'_3\tau_3} \left[g^{-1}\right]^{l'_4l_4}_{\tau'_4\tau_4}
\end{align}
Here, $\av{\ldots}$ stands for the average with respect to the partition function of the local impurity problem~\eqref{eq:action_imp_app}.
After integrating out original fermionic variables, we perform a transformation from unphysical bosonic fields $\phi^{\varsigma}_{ll'}$ to physical variables $\bar{\rho}^{\varsigma}_{ll'}$ that describe fluctuations of charge and spin densities.
To this aim we make a Hubbard-Stratonovich transformation for the quadratic in $\hat{\phi}^{\,\varsigma}_{ll'}$ variables term that enters the last line of Eq.~\eqref{eq:action_pp4}, which is the only term in the action that contains bosonic source fields $j^{\,\varsigma}$ 
\begin{align}
&\exp\left\{\frac12\int_{0}^{\beta} d\tau \sum_{ij,\varsigma,\{l\}} \left(\phi^{\varsigma}_{i\tau{}l_1l_2} - \sum_{\varsigma'}{\cal A}^{\varsigma'}_{i\tau} [{\cal U}_{i\tau}^{-1}]^{\varsigma'\varsigma} \delta^{\phantom{\dagger}}_{l_1l_2} - j^{\,\varsigma}_{i\tau{}l_1l_2}\right) \left[\left(V^{\varsigma}\right)^{-1}\right]^{ij}_{l_1l_2,l_3l_4} \left(\phi^{\varsigma}_{j\tau{}l_4l_3} - \sum_{\varsigma''}{\cal A}^{\varsigma''}_{j\tau} [{\cal U}_{j\tau}^{-1}]^{\varsigma''\varsigma} \delta^{\phantom{\dagger}}_{l_4l_3} - j^{\,\varsigma}_{j\tau{}l_4l_3}\right) \right\} 
= \notag\\
&\left[-{\cal D}^{-1}_{\phi}\right]
\int D[\bar{\rho}^{\varsigma}] \exp\left\{- \int_{0}^{\beta} d\tau \sum_{ij,\varsigma} \left( \frac12 \sum_{\{l\}}\bar{\rho}^{\varsigma}_{i\tau{l_1l_2}} V^{\varsigma,\,ij}_{l_1l_2l_3l_4} \, \bar{\rho}^{\varsigma}_{j\tau{}l_4l_3} - \sum_{ll'}\left(\phi^{\varsigma}_{i\tau{}ll'} - \sum_{\varsigma'}{\cal A}^{\varsigma'}_{i\tau} [{\cal U}_{i\tau}^{-1}]^{\varsigma'\varsigma}\delta^{\phantom{\dagger}}_{ll'} - j^{\,\varsigma}_{i\tau{}ll'}\right) \bar{\rho}^{\varsigma}_{i\tau{}l'l} \right)\right\}
\label{eq:HSb2_app}
\end{align}
After that the action takes the following form
\begin{align} 
{\cal S}
= 
&-\int_{0}^{\beta} \{d\tau_{i}\} \Tr_{\sigma} \sum_{ij,\{l\}} 
f^{*}_{i\tau_1l_1} \Bigg\{ \left[g^{-1}\right]^{l_1l_2}_{\tau_1\tau_2}
\left[\tilde{\varepsilon}^{-1}\right]^{\tau_2\tau_3,l_2l_3}_{ij,\sigma\sigma'}\left[g^{-1}\right]^{l_3l_4}_{\tau_3\tau_4}
- \delta_{ij}^{\phantom{*}}\left[g^{-1}\right]^{l_1l_2}_{\tau_1\tau_2} R^{\phantom{\dagger}}_{i\tau_2} g^{l_2l_3}_{\tau_2\tau_3} R^{\dagger}_{i\tau_3} \left[g^{-1}\right]^{l_3l_4}_{\tau_3\tau_4} \Bigg\} f^{\phantom{*}}_{j\tau_4l_4} \notag\\
&+ \frac12 \int_{0}^{\beta} d\tau\,\sum_{ij,\varsigma,\{l\}} \bar{\rho}^{\varsigma}_{i\tau{}l_1l_2} V^{\varsigma,\,ij}_{l_1l_2l_3l_4} \, \bar{\rho}^{\varsigma}_{j\tau{}l_4l_3}
+ \frac12\iint_{0}^{\beta} d\tau \, d\tau' \sum_{i,\{\varsigma\},\{l\}} \phi^{\varsigma_1}_{i\tau{}l_1l_2} {\cal U}^{\varsigma_1\varsigma_2}_{i\tau} \chi^{\varsigma_2,\,\tau\tau'}_{l_1l_2l_3l_4} [{\cal U}_{i\tau'}^{-1}]^{\varsigma_2\varsigma_3} \phi^{\varsigma_3}_{i\tau'l_4l_3} \notag\\
&+ \int_{0}^{\beta} d\tau \, \Bigg\{\sum_{i,mm',l}{\cal A}^{m}_{i\tau} [{\cal U}_{i\tau}^{-1}]^{mm'} \bar{\rho}^{m'}_{i\tau{}ll}
- \sum_{i,\varsigma,ll'} 
\left[\phi^{\varsigma}_{i\tau{}ll'} - j^{\,\varsigma}_{i\tau{}ll'} \right]\bar{\rho}^{\varsigma}_{i\tau{}l'l} 
\Bigg\}
+\tilde{\cal F}[f,\phi]
\label{eq:action_ppp}
\end{align}
We find that the source field $j^{\,\varsigma}_{i\tau}$ enters the problem only as a multiplier of the new field $\bar{\rho}^{\varsigma}_{i\tau}$. 
This means that the introduced field $\bar{\rho}^{\varsigma}_{i\tau}$ is a correct variable that describes fluctuations of charge and spin densities as discussed in the main text. 
Further we omit bosonic sources $j^{\,\varsigma}_{i\tau}$ and bars over the bosonic field $\rho^{\varsigma}_{i\tau}$.
Finally, bosonic fields $\phi^{\varsigma}$ can be integrated out as 
\begin{align}
&\int D[\phi^{\varsigma}] \exp\Bigg\{
-\frac12\iint_{0}^{\beta} d\tau \, d\tau' \sum_{i,\{\varsigma\},\{l\}} \phi^{\varsigma_1}_{i\tau{}l_1l_2} {\cal U}^{\varsigma_1\varsigma_2}_{i\tau} \chi^{\varsigma_2,\,\tau\tau'}_{l_1l_2l_3l_4} [{\cal U}_{i\tau'}^{-1}]^{\varsigma_2\varsigma_3} \phi^{\varsigma_3}_{i\tau'{}l_4l_3} \notag\\
&\hspace{2.15cm}+ \int_{0}^{\beta} d\tau_4 \sum_{i,\varsigma',l_5l_6}\left(\rho^{\varsigma'}_{i\tau_4l_5l_6} 
- \iiint_{0}^{\beta} d\tau_1 \, d\tau_2 \, d\tau_3 \Tr_{\sigma} \sum_{\varsigma, l_1l_2l_3l_4} f^{*}_{i\tau_1l_1}
R^{\phantom{\dagger}}_{i\tau_1}\sigma^{\varsigma}R^{\dagger}_{i\tau_2}
f^{\phantom{*}}_{i\tau_2l_2} \Lambda^{\varsigma,\,\tau_1\tau_2\tau_3}_{l_1l_2;l_3l_4} \chi^{\varsigma,\,\tau_3\tau_4}_{l_3l_4l_5l_6} [{\cal U}_{i\tau_4}^{-1}]^{\varsigma\varsigma'} \right) \, \phi^{\varsigma'}_{i\tau_4l_6l_5} \Bigg\} = \notag\\
&{\cal Z}_{\phi} \times 
\exp\Bigg\{
\frac12\iint_{0}^{\beta} d\tau \, d\tau' \sum_{i,\{\varsigma\},\{l\}} \rho^{\varsigma_1}_{i\tau{}l_1l_2} {\cal U}^{\varsigma_1\varsigma_2}_{i\tau} \left[\left(\chi^{\varsigma_2}\right)^{-1}\right]^{\tau\tau'}_{l_1l_2,l_3l_4} [{\cal U}^{-1}_{i\tau'}]^{\varsigma_2\varsigma_3} \rho^{\varsigma_3}_{i\tau'l_4l_3} \notag\\
&\hspace{1.5cm}- \int_{0}^{\beta} \{d\tau\} \Tr_{\sigma} \sum_{i,\varsigma\varsigma',\{l\}} f^{*}_{i\tau_1l_1}
R^{\phantom{\dagger}}_{i\tau_1}\sigma^{\varsigma}R^{\dagger}_{i\tau_2}
f^{\phantom{*}}_{i\tau_2l_2} \Lambda^{\varsigma,\,\tau_1\tau_2\tau_3}_{l_1l_2;l_3l_4} [{\cal U}^{-1}_{i\tau_3}]^{\varsigma\varsigma'} \rho^{\varsigma'}_{i\tau_3l_4l_3}
 \notag\\
&\hspace{1.5cm}+\frac12 \int_{0}^{\beta} \{d\tau\} \Tr_{\sigma} \sum_{i,\varsigma,\{l\}} f^{*}_{i\tau_1l_1}
R^{\phantom{\dagger}}_{i\tau_1}\sigma^{\varsigma}R^{\dagger}_{i\tau_2}
f^{\phantom{*}}_{i\tau_2l_2} \Lambda^{\varsigma,\,\tau_1\tau_2\tau_3}_{l_1l_2;l_3l_4} \chi^{\varsigma,\,\tau_3\tau_4}_{l_3l_4l_5l_6} \Lambda^{\varsigma,\,\tau_6\tau_5\tau_4}_{l_8l_7;l_6l_5} f^{*}_{i\tau_6l_8}
R^{\phantom{*}}_{i\tau_6}\sigma^{\varsigma}R^{\dagger}_{i\tau_5}
f^{\phantom{*}}_{i\tau_5l_7}
\Bigg\}
\label{eq:Integration_phi_app}
\end{align}
where ${\cal Z}_{\phi}$ is a partition function of the Gaussian part of this bosonic integral. 
The four-point interaction that appears in the last line of this expression approximately cancels the exact four-point vertex function $\Gamma$ (second term in Eq.~\eqref{eq:lowestint}) as discussed in Ref.~\cite{PhysRevLett.121.037204}.
As the result we get
\begin{align} 
{\cal S}
= 
&-\int_{0}^{\beta} \{d\tau_{i}\} \Tr_{\sigma} \sum_{ij,\{l\}} 
f^{*}_{i\tau_1l_1} \Bigg\{ \left[g^{-1}\right]^{l_1l_2}_{\tau_1\tau_2}
\left[\tilde{\varepsilon}^{-1}\right]^{\tau_2\tau_3,l_2l_3}_{ij,\sigma\sigma'}\left[g^{-1}\right]^{l_3l_4}_{\tau_3\tau_4}
- \delta_{ij}^{\phantom{*}}\left[g^{-1}\right]^{l_1l_2}_{\tau_1\tau_2} R^{\phantom{\dagger}}_{i\tau_2} g^{l_2l_3}_{\tau_2\tau_3} R^{\dagger}_{i\tau_3} \left[g^{-1}\right]^{l_3l_4}_{\tau_3\tau_4} \Bigg\} f^{\phantom{*}}_{j\tau_4l_4}  \notag\\
&+ \frac12 \int_{0}^{\beta} d\tau\,\sum_{ij,\varsigma,\{l\}} \rho^{\varsigma}_{i\tau{}l_1l_2} V^{\varsigma,\,ij}_{l_1l_2l_3l_4} \, \rho^{\varsigma}_{j\tau{}l_4l_3}
- \frac12\iint_{0}^{\beta} d\tau \, d\tau' \sum_{i,\{\varsigma\},\{l\}} \rho^{\varsigma_1}_{i\tau{}l_1l_2} {\cal U}^{\varsigma_1\varsigma_2}_{i\tau} \left[\left(\chi^{\varsigma_2}\right)^{-1}\right]^{\tau\tau'}_{l_1l_2,l_3l_4} [{\cal U}_{i\tau'}^{-1}]^{\varsigma_2\varsigma_3} \rho^{\varsigma_3}_{i\tau'l_4l_3} \notag\\
&+ \int_{0}^{\beta} \{d\tau\} \Tr_{\sigma} \sum_{i,\varsigma\varsigma',\{l\}} f^{*}_{i\tau_1l_1}
R^{\phantom{\dagger}}_{i\tau_1}\sigma^{\varsigma}R^{\dagger}_{i\tau_2}
f^{\phantom{*}}_{i\tau_2l_2} \Lambda_{l_1l_2;l_3l_4}^{\varsigma,\,\tau_1\tau_2\tau_3} [{\cal U}^{-1}_{i\tau_3}]^{\varsigma\varsigma'} \rho^{\varsigma'}_{i\tau_3l_4l_3}
+\int_{0}^{\beta} d\tau \sum_{i,ss',l}{\cal A}^{s}_{i\tau} [{\cal U}_{i\tau}^{-1}]^{ss'} \rho^{s'}_{i\tau{}ll}
\label{eq:action_ppp3}
\end{align}
We assume that the multi-orbital system that exhibits a well-developed magnetic moment is characterised by a strong Hund's exchange coupling that orders spins of electrons at each orbital in the same direction. 
This allows one to decouple the orbital and spin degrees of freedom rewriting the vector spin field as $\rho^{s}_{i\tau{}ll'} = M^{\phantom{*}}_{i\tau{}ll'} e^{s}_{i\tau}$. Here, $M_{i\tau{}ll'}$ is a scalar field that can be associated with the orbitally-resolved value of the local magnetic moment. In its turn, $\vec{e}_{i\tau}$ is the unit vector that points in the direction of the local moment on the lattice site $i$ at the time $\tau$. 
At this step we set the direction of the $z$-axis defined by the rotation matrices~\eqref{eq:Rmatr_app} equal to the direction of the local magnetic moment. 
Then, using the definition of ${\cal U}^{ss'}$ we get
\begin{align}
\sum_{s}\rho^{s}_{i\tau{}ll'} {\cal U}^{ss'}_{i\tau} 
&= \sum_{s} M^{\phantom{*}}_{i\tau{}ll'} e^{s}_{i\tau} {\cal U}^{ss'}_{i\tau}
= M^{\phantom{*}}_{i\tau{}ll'} \delta^{\phantom{*}}_{z,s'} 
\end{align}
The validity of this approximation will be discussed in Appendix~\ref{app:SPA} in order not to confuse the reader of this extremely technically loaded appendix even more.
In addition, we use that the 
local Green's function ${g^{l_1l_2}_{\tau_1 \tau_2}}$ dominates at ${\tau_1 \approx \tau_2}$ and is diagonal in the spin space. This allows one to write that
\begin{align}
\iint^{\beta}_{0} d\tau_2 \, d\tau_3 \Tr_{\sigma} \sum_{l_2l_3}
f^{*}_{i\tau_1l_1} \left[g^{-1}\right]^{l_1l_2}_{\tau_1\tau_2} R^{\phantom{\dagger}}_{i\tau_2} g^{l_2l_3}_{\tau_2\tau_3} R^{\dagger}_{i\tau_3} \left[g^{-1}\right]^{l_3l_4}_{\tau_3\tau_4} f^{\phantom{*}}_{i\tau_4l_4} 
&\simeq 
\int^{\beta}_{0} d\tau_3 \Tr_{\sigma} \sum_{l_3} \delta^{\phantom{*}}_{\tau_1\tau_3} \delta^{\phantom{*}}_{l_1l_3}
f^{*}_{i\tau_1l_1} R^{\phantom{\dagger}}_{i\tau_3} R^{\dagger}_{i\tau_3} \left[g^{-1}\right]^{l_3l_4}_{\tau_3\tau_4} f^{\phantom{*}}_{i\tau_4l_4} \notag\\
&= \Tr_{\sigma} 
f^{*}_{i\tau_1l_1} \left[g^{-1}\right]^{l_1l_4}_{\tau_1\tau_4} f^{\phantom{*}}_{i\tau_4l_4} 
\end{align}
Using the same argument, the three-point vertex can be transformed as
\begin{gather}
\Tr_{\sigma} \sum_{ss',\{l\}} f^{*}_{i\tau_1l_1}
R^{\phantom{\dagger}}_{i\tau_1}\sigma^{s}R^{\dagger}_{i\tau_2}
f^{\phantom{*}}_{i\tau_2l_2} \Lambda_{l_1l_2l_3l_4}^{\hspace{-0.05cm}s,\,\tau_1\tau_2\tau_3} [{\cal U}_{i\tau_3}^{-1}]^{ss'} \rho^{s'}_{i\tau_3l_4l_3}
= \Tr_{\sigma} \sum_{\{l\}} f^{*}_{i\tau_1l_1}
R^{\phantom{\dagger}}_{i\tau_1}\sigma^{z}R^{\dagger}_{i\tau_2}
f^{\phantom{*}}_{i\tau_2l_2} \Lambda^{\hspace{-0.05cm}z,\,\tau_1\tau_2\tau_3}_{l_1l_2;l_3l_4}
M^{\phantom{*}}_{i\tau_3l_4l_3} \simeq \notag\\
\sum_{s,\sigma\sigma',\{l\}} f^{*}_{i\tau_1\sigma{}l_1}
\sigma^{s}_{\sigma\sigma'}
f^{\phantom{*}}_{i\tau_2\sigma'l_2} \Lambda^{\hspace{-0.05cm}z,\,\tau_1\tau_2\tau_3}_{l_1l_2;l_3l_4}
M^{\phantom{*}}_{i\tau_3l_4l_3} e^{s}_{i\tau_3} =
\sum_{s,\sigma\sigma'\{l\}} f^{*}_{i\tau_1\sigma{}l_1}
\sigma^{s}_{\sigma\sigma'}
f^{\phantom{*}}_{i\tau_2\sigma'l_2} \Lambda^{\hspace{-0.05cm}s,\,\tau_1\tau_2\tau_3}_{l_1l_2;l_3l_4}
\rho^{s}_{i\tau_3l_4l_3}
\end{gather}
After that, one gets the fermion-boson action
\begin{align} 
{\cal S}
= 
&-\iint_{0}^{\beta} d\tau \, d\tau' \sum_{ij,\sigma\sigma'}  
f^{*}_{i\tau\sigma{}l} \left[\tilde{\cal G}^{-1} \right]^{\tau\tau',ll'}_{ij,\sigma\sigma'}
f^{\phantom{*}}_{j\tau'\sigma'l'}
+ \int_{0}^{\beta} \{d\tau\} \sum_{i,\varsigma,\sigma\sigma',\{l\}} f^{*}_{i\tau_1\sigma{}l_1}\sigma^{\varsigma}_{\sigma\sigma'}
f^{\phantom{*}}_{i\tau_2\sigma'l_2} \Lambda^{\varsigma,\,\tau_1\tau_2\tau_3}_{l_1l_2;l_3l_4} \rho^{\varsigma}_{i\tau_3l_4l_3}
+ \frac12 \int_{0}^{\beta} d\tau\,\sum_{ij,\varsigma,\{l\}} \rho^{\varsigma}_{i\tau{}l_1l_2} V^{\varsigma,\,ij}_{l_1l_2l_3l_4}  \rho^{\varsigma}_{j\tau{}l_4l_3} \notag\\
&- \frac12 \iint_{0}^{\beta} d\tau\,d\tau' \sum_{i,\{l\}} \rho^{c}_{i\tau{}l_1l_2} \left[\left(\chi^{c}\right)^{-1}\right]^{\tau\tau'}_{l_1l_2,l_3l_4} \rho^{c}_{i\tau'l_4l_3}
- \frac12 \iint_{0}^{\beta} d\tau\,d\tau' \sum_{i,\{l\}} M^{\phantom{*}}_{i\tau{}l_1l_2} \left[\left(\chi^{z}\right)^{-1}\right]^{\tau\tau'}_{l_1l_2,l_3l_4} M^{\phantom{*}}_{i\tau'l_4l_3}
+ \int_{0}^{\beta} d\tau \sum_{i} {\cal A}^{z}_{i\tau} {\cal M}^{\phantom{*}}_{i\tau}
\label{eq:S_dual_app}
\end{align}
where ${{\cal A}^{z}_{i\tau} = \frac12 i \dot{\varphi} \, (1-\cos\theta)}$ is the Berry phase, ${{\cal M}_{i\tau} = \sum_{l}M^{\phantom{*}}_{i\tau{}ll}}$ is the modulus of the total local magnetic moment, and $\tilde{\cal G}$ is the non-local part of the DMFT Green's function
\begin{align}
\tilde{\cal G}^{\tau\tau'll'}_{ij\sigma\sigma'} = G^{\tau\tau'll'}_{ij\sigma\sigma'} - \delta^{\phantom{*}}_{ij}\delta^{\phantom{*}}_{\sigma\sigma'} g^{ll'}_{\tau\tau'}
\label{eq:G_dual_app}
\end{align}
~

\twocolumngrid
\subsubsection*{Summary of performed transformations}

To guide the reader through a technical derivation presented in this section and to clarify the meaning of the introduced auxiliary fields, let us make a summary of the performed transformations.
We start with the fermionic action given by Eq.~\eqref{eq:action_latt_app}. 
The corresponding partition function ${{\cal Z} = \int D[c^{*},c] \, e^{-{\cal S}[c^{(*)}]}}$ for this initial fermionic action ${\cal S}[c^{(*)}]$ contains the integral over original fermionic variables $c^{(*)}$.
Then, we introduce an exactly (numerically) solvable local reference system ${\cal S}_{\rm imp}[c^{(*)}]$~\eqref{eq:action_imp_app} that further will be integrated out in order to account for local correlation effects exactly.
We note that the reference system and the remaining part of the action ${{\cal S}_{\rm rem}[c^{(*)}] = {\cal S}[c^{(*)}] - \sum_{i} {\cal S}_{\rm imp}[c^{(*)}]}$ are written in terms of the same variables $c^{(*)}$.
Therefore, the reference system can be integrated out only after changing variables for the remaining part of the action.
This change of variables can be done using two Hubbard-Stratonovish transformations~\eqref{eq:HSf1_app} and~\eqref{eq:HSb1_app}, which introduces new fermionic $f^{(*)}$ and bosonic $\phi^{\varsigma}$ fields in the theory.
After that, the partition function becomes ${{\cal Z} = \int D[c^{*},c] D[f^{*},f] D[\phi^{\varsigma}] \, e^{-{\cal S}[c^{(*)}, f^{(*)}, \phi^{\varsigma}]}}$.
To simplify expressions, hereinafter the partition function is given up to insignificant factors, e.g. determinants, that are explicitly specified above.
We note that new fermionic $f^{(*)}$ and bosonic $\phi^{\varsigma}$ fields have no physical meaning and are introduced only to enable integrating the reference system out.
Before performing this integration, we make a transformation of fermionic variables $c^{(*)}$ to a rotating frame using a rotation matrix~\eqref{eq:Rmatr_app}.
This allows us to introduce polar angles $\Omega_{R}=\{\theta_{i\tau},\varphi_{i\tau}\}$ that further will be associated with the direction of the bosonic field that describes fluctuations of the local magnetic moment.
Similar to Refs.~\cite{PhysRevLett.65.2462, PhysRevB.43.3790, doi:10.1142/S0217979200002430, DUPUIS2001617}, we introduce path integration over the angles $\Omega_{R}$ to maintain rotational invariance.
This results in the following partition function ${{\cal Z} = \int D[c^{*},c] D[f^{*},f] D[\phi^{\varsigma}] D[\Omega_{R}] \, e^{-{\cal S}[c^{(*)}, f^{(*)}, \phi^{\varsigma}, \Omega_R]}}$, where by $D[\Omega_R] = \prod \sin\theta \,d\theta\, d\phi/4\pi$ we mean the usual differential in spherical coordinates. 
Now, one can integrate out original fermionic variables $c^{(*)}$~\eqref{eq:Integrating_S_imp_app}, which results in the  partition function ${{\cal Z} = \int D[f^{*},f] D[\phi^{\varsigma}] D[\Omega_R] \, e^{-{\cal S}[f^{(*)}, \phi^{\varsigma},\Omega_R]}}$ 
that is not specified in the text. 
Further, we make another Hubbard-Stratonovich transformation~\eqref{eq:HSb2_app} in order to go from an auxiliary bosonic field $\phi^{\varsigma}$ to a physical bosonic field $\bar{\rho}^{\varsigma}$ that describes fluctuations of charge and spin densities. 
The partition function for the problem~\eqref{eq:action_ppp} transforms to 
${{\cal Z} = \int D[f^{*},f] D[\phi^{\varsigma}] D[\bar{\rho}^{\varsigma}] D[\Omega_R] \, e^{-{\cal S}[f^{(*)}, \phi^{\varsigma}, \bar{\rho}^{\varsigma}, \Omega_R]}}$. 
Finally, auxiliary bosonic fields $\phi^{\varsigma}$ are integrated out~\eqref{eq:Integration_phi_app}, which results in the partition function
${{\cal Z} = \int D[f^{*},f] D[\bar{\rho}^{\varsigma}] D[\Omega_R] \, e^{-{\cal S}[f^{(*)}, \bar{\rho}^{\varsigma}, \Omega_R]}}$ for the
fermion-boson action~\eqref{eq:action_ppp3}. 
At the last step we notice that under certain conditions (see Appendix~\ref{app:SPA}) the path integral over the direction of the spin part bosonic field $\bar{\rho}^{s}$ can be taken in the saddle point approximation that allows to associate the rotation angles $\Omega_R$ with the direction of the aforementioned field.
The latter is a vector field with three components $\bar{\rho}^{x}$, $\bar{\rho}^{y}$, and $\bar{\rho}^{z}$. 
Therefore, one can make a transformation from Cartesian to spherical coordinate system and identically rewrite this field as ${\rho^{s} = M e^{s}}$.
Here, $M_{i\tau}$ is a scalar field that can be associated with the modulus of the local magnetic moment, and $\vec{e}_{i\tau}$ is the unit vector that points in the direction of the local moment on the lattice site $i$ at the time $\tau$.
Let us assume that the direction of the unit vector $\vec{e}_{i\tau}$ is defined by a set of polar angles ${\Omega_M=\{\theta'_{i\tau},\varphi'_{i\tau}\}}$. 
Then, the integral in the partition function ${{\cal Z} = \int D[f^{*},f] D[\bar{\rho}^{\varsigma}] D[\Omega_{R}] \, e^{-{\cal S}[f^{(*)}, \bar{\rho}^{\varsigma}, \Omega_R]}}$ can also be written as ${{\cal Z} = \int D[f^{*},f] D[\bar{\rho}^{c}] D[\Omega_R] D[M, \Omega_M] \, e^{-{\cal S}[f^{(*)}, \bar{\rho}^{c}, \Omega_R, M, \Omega_M]}}$, where $D[M,\Omega_M] = \prod M^2 \sin\theta' dM d\theta' d\phi'/4\pi$.
The saddle point approximation discussed in Appendix~\ref{app:SPA} allows one to equate the angles ${\Omega_R=\Omega_M}$ that from now on define the direction of the unit vector field $\vec{e}_{i\tau}$ and consequently of the local magnetic moment. After that, the partition function of the problem reduces to its final form ${{\cal Z} = \int D[f^{*},f] D[\bar{\rho}^{c}] D[M, \Omega_M] \, e^{-{\cal S}[f^{(*)}, \bar{\rho}^{c}, M, \Omega_M]}}$.

\section{Validity of the saddle point approximation.}
\label{app:SPA}

In this appendix we argue that under certain conditions the path integral over $\Omega_M$ effectively acts as a delta-function setting all fields $e^s_{i\tau}$ in the action (\ref{eq:S3}) effectively equal to the unit vectors in the directions of the $z$-axes of the rotating reference frames described by angles $\Omega_R$. 
As discussed in the main text, the 3rd and the 6th terms in Eq.~\eqref{eq:S3} account for the dynamics of the local magnetic moment. 
Other terms either account for charge dynamics and thus do not depend on $\Omega_M$ or account for non-local exchange interaction between the local moments  and thus are irrelevant for the internal dynamics of the local moment. 
They are however important to derive the equation of motion of the moments the solution of which is essential to verify whether the approximation made is consistent (see below). So, consider the expression for the partition function:
\begin{align}
    {\cal Z} &= \int D[\Omega_M,\dots] \, e^{-{\cal S}_r} \,  \times \notag\\ &\times \exp{\left\{-\frac12\iint_{0}^{\beta} d\tau\,d\tau' \sum_{i,\{\varsigma\}} \bar{\rho}^{\,\varsigma_1}_{i\tau} {\cal U}^{\varsigma_1\varsigma_2}_{i\tau} \left[\chi^{\varsigma_2}\right]^{-1}_{\tau\tau'} [{\cal U}_{i\tau'}^{-1}]^{\varsigma_2\varsigma_3} \bar{\rho}^{\,\varsigma_3}_{i\tau'}\right\}}
\label{eq:B1}
\end{align}
Here ${\cal S}_r$ stands for all other terms in the action except the 3rd term in Eq.~\eqref{eq:S3} and ``$\dots$'' stands for all other fields except $\Omega_M$. The saddle-point approximation is well-justified if the integrand is a product of a slow function of the integration ($e^{-{\cal S}_r}$) variable and a sharp-peaked function of that variable (the exponent of the 3rd term). Here we will discuss the conditions under which this is a good approximation.

Consider first the 3rd term.
It couples the spin bosonic field in the rotating reference frame ${\rho'_{i\tau}=[{\cal U}_{i\tau}^{-1}] \, \bar{\rho}_{i\tau}}$ at the site $i$ and time $\tau$ to that very same field $\rho'_{i\tau'}$ at a different point of time $\tau'$. 
As we consider the paramagnetic phase, the coupling constant between these two fields, which happens to be the inverse of the local magnetic susceptibility, does not depend on the direction in space. 
As also is discussed in the main text and is shown in Fig.~\ref{fig:chi_loc} the susceptibility decays rapidly with frequency meaning that its dependence on ${\tau-\tau'}$ is very slow.
Note that the value of the susceptibility defined as ${\chi^{\varsigma}_{\tau\tau'} = - \langle \rho^{\varsigma}_{\tau} \rho^{\varsigma}_{\tau'} \rangle}$ is negative, and the Fig.~\ref{fig:chi_loc} shows the absolute value of it.
As the scalar product of two unit vectors is maximized when the vectors are parallel, the exponent of the 3rd term is maximized if the direction of the vector bosonic field $\rho'_{i\tau}$ at site $i$ is a constant in the rotating reference frame at the same site. 
The absolute value of the inverse susceptibility must be sufficiently large for the saddle-point approximation to be valid. We will get back to that at the end of this Appendix. 
If we choose the initial conditions in our path integration by setting the directions of the $z$-axis of the rotating frame and the local magnetic moment equal at time ${\tau=0}$, we can conclude that the saddle trajectory is indeed the one where ${\Omega_R=\Omega_M}$.  
Having established this, we can move on with deriving the equations of motion for the moments in the lattice as is done in the main text, and the solution of those equations would then be the saddle trajectory for the angles $\Omega_R$ and consequently also for $\Omega_M$. An important note is that ${\cal S}_r$ must be small along the saddle trajectory for the approximation to be valid. Let us take a closer look at this condition.

Consider the 6th term in Eq.~\eqref{eq:S3} (the Berry phase).
\begin{equation}
    {\cal S}_6=i\int_0^{\beta} d\tau \sum_{is}\mathcal{A}_{i\tau}^s \, \rho'^s_{i\tau}
\end{equation}
$\mathcal{A}$ here depends on $\Omega_R$ only and is proportional to its time derivative.
As long as the spin dynamics it describes is slow {\it i.e.} adiabatic approximation is applicable and the local moment is well-defined, this term remains small compared to the 3rd term discussed above and thus does not affect the saddle trajectory. 
The small parameter here is ${\sim J/U}$, with $J$ being the characteristic value of the exchange interaction Eq.~\eqref{eq:J}. 
When the solution of the equations of motion becomes such that the time derivatives of the angles in the Berry phase term become large enough, this purely imaginary term leads to suppression of the weight of the corresponding trajectories in the path integral due to quantum fluctuations. In this sense the pinning of the rotating reference frame to the direction of the local moment can be justified only if the local moment is indeed well-defined. The internal consistency of the saddle point approximation is thus ensured if after considering the exchange interaction arising from other terms in Eq.~\eqref{eq:S3} the resultant solution of the LLG equation~\eqref{eq:eq_motion} is slow. 

Another important aspect becomes apparent if we estimate the value of the 3rd term in Eq.~\eqref{eq:S3}. This term is proportional to the scalar product of the unity vectors describing the directions of the vector field in the rotating frame at two different points of time (and this scalar product is equal to 1 if that field is pinned to the rotating frame). The proportionality coefficient is ${\approx M^2/2T\chi^{s}_{\omega=0}}$. In order for the saddle point approximation to work this coefficient must be large compared to unity. Our estimates show that the value of the local moment in the considered single-band case must be at least ${\sim3}$ for this to be valid. This basically means that the saddle point approximation works only in the multi-orbital case, where the large value of the magnetic moment is provided by a strong Hund's coupling. This actually makes sense as only in this case we can expect the Berry-phase term to have its classical form and consequently the dynamics of the local moments to be described by the classical Heisenberg model with exchange interactions. If the value of the local moment becomes small, the quantum fluctuations play an increasingly important role and we switch to a regime, where the local moment might still be well-defined in the sense that the criterion for the formation of the local moment derived in Section~\ref{sec:local_moment} still makes sense, but the dynamics of those moments are in no sense classical.

\onecolumngrid

\section{Local free energy}
\label{app:local}

In this section we derive the local free energy related to the local magnetic moment.
The total local free energy of the system is given by the reference system~\eqref{eq:action_imp_app}.
However, as pointed out in the main text, this local problem describes all local correlations including the effect of itinerant electrons. 
In order to isolate the part of the energy related to the local magnetic moment the electronic contribution should be subtracted from the total local energy. 
To this effect we start with the action~\eqref{eq:S_HS_app} that is reminiscent of the ${s\text{-}d}$ model, since it describes a set of localized impurities~\eqref{eq:action_imp_app} that are coupled to each other via itinerant electrons $f^{(*)}$. 
If one excludes non-local contributions, which also contain the dispersion of the itinerant $f^{(*)}$ electrons, from this action, one gets the following effective local action
\begin{align} 
{\cal S}_{\rm loc}
= \iint_{0}^{\beta} d\tau\, d\tau' \sum_{\sigma,ll'}\left( c^{*}_{\tau\sigma{}l}\left[g^{-1}\right]^{ll'}_{\tau\tau'}f^{\phantom{*}}_{\tau'\sigma{}l'} + f^{*}_{\tau\sigma{}l}\left[g^{-1}\right]^{ll'}_{\tau\tau'}c^{\phantom{*}}_{\tau'\sigma{}l'}
\right)
+\int_{0}^{\beta} d\tau \sum_{i,\varsigma,ll'} \phi^{\varsigma}_{\tau{}ll'} \rho^{\varsigma}_{\tau{}l'l}
+ {\cal S}_{\rm imp}
\label{eq:S_loc2_app}
\end{align}
One finds that fields $\hat{f}^{(*)} = f^{(*)}g^{-1}$ and $\phi^{\varsigma}$ play a role of source fields for electronic $c^{(*)}$ and composite variables $\rho^{\varsigma}$, respectively. 
Therefore, if now we integrate out original fermionic degrees of freedom by expanding a partition function in powers of these effective source fields, as it was done in Eq.~\eqref{eq:Integrating_S_imp_app}, we get the following action
\begin{align}
{\cal S}_{\rm loc} = -\ln{\cal Z}_{\rm imp} - \iint_{0}^{\beta} d\tau \, d\tau' \sum_{\sigma,ll'}
f^{*}_{\tau\sigma{}l} \left[- \, g^{-1}\right]^{ll'}_{\tau\tau'} f^{\phantom{*}}_{\tau'l'} 
- \frac12 \iint^{\beta}_{0} d\tau \, d\tau' \sum_{\varsigma, \{l\}} \phi^{\varsigma}_{\tau{}l_1l_2} \left[-\,\chi^{\varsigma,\,\tau\tau'}_{l_1l_2l_3l_4}\right] \phi^{\varsigma}_{\tau'l_4l_3}
+\tilde{\cal F}[f,\phi]
\label{eq:S_loc3_app}
\end{align}
where the first term gives the total local energy of the system, and the interaction part $\tilde{\cal F}[f,\phi]$ defined in Appendix~\ref{app:HS} contains all possible exact fermion-fermion, fermion-boson, and boson-boson vertex functions of the local reference problem~\eqref{eq:action_imp_app}. 
Importantly, we find that in the obtained action the bare fermionic Green's function for fields $f^{(*)}$ coincides with the exact Green's function of the impurity problem $g$. 
Integrating out $f^{(*)}$ variables in this expression generates diagrammatic corrections to boson-boson interactions that are constructed from the exact local fermion-boson vertices connected by the exact Green's functions of the impurity problem.
We observe that the generated itinerant contributions become subtracted from the total boson-boson contributions to the full local energy of the system.
Thus, after integrating $f^{(*)}$ fields out the remaining part of the local action describes the part of the local energy related to bosonic fields $\phi^{\varsigma}$ only. 
Our aim is to derive the corresponding action for physical bosonic variables $\rho^{\varsigma}$.
Therefore, we perform the same procedure for the action~\eqref{eq:action_ppp3}, namely we exclude the non-local (itinerant) terms from this action and integrate fermionic fields as
\begin{gather}
\int D[f^{*},f]\,\exp\left\{ 
- \iint_{0}^{\beta} d\tau\,d\tau' \sum_{\sigma\sigma',ll'}
f^{*}_{\tau\sigma{}l} \left[\left[g^{-1}\right]_{\tau\tau'}^{ll'}\delta^{\phantom{*}}_{\sigma\sigma'} + \int^{\beta}_{0} d\tau''\sum_{\varsigma,l''l'''} \sigma^{\varsigma}_{\sigma\sigma'}\Lambda^{\varsigma,\,\tau\tau'\tau''}_{ll';l''l'''} \rho^{\varsigma}_{i\tau''l''l'''} \right] f^{\phantom{*}}_{\tau'\sigma'l'}\right\} = \notag\\
{\rm det}\left[\left[g^{-1}\right]_{\tau\tau'}^{ll'}\delta^{\phantom{*}}_{\sigma\sigma'} + \int^{\beta}_{0} d\tau''\sum_{\varsigma,l''l'''} \sigma^{\varsigma}_{\sigma\sigma'}\Lambda^{\varsigma,\,\tau\tau'\tau''}_{ll';l''l'''} \rho^{\varsigma}_{i\tau''l''l'''} \right]
\end{gather}
Note that the rotation of the quantization axis of fermions in the case of a local problem is a trivial procedure. For this reason, we do no perform the transformation of fermionic variables to a rotating frame~\eqref{eq:Rmatr_app} in this section.
Then, the total local action that describes the physics of the local magnetic moment reads
\begin{align} 
{\cal S}_{\rm loc}
= &-\Tr\ln\left[\left[g^{-1}\right]_{\tau\tau'}^{ll'}\delta^{\phantom{*}}_{\sigma\sigma'} + \int^{\beta}_{0} d\tau''\sum_{\varsigma,l''l'''} \sigma^{\varsigma}_{\sigma\sigma'}\Lambda^{\varsigma,\,\tau\tau'\tau''}_{ll';l''l'''} \rho^{\varsigma}_{\tau''l''l'''} \right]
- \frac12 \iint_{0}^{\beta} d\tau\,d\tau' \sum_{\varsigma,\{l\}} \rho^{\varsigma}_{i\tau{}l_1l_2} \left[\left(\chi^{\varsigma}\right)^{-1}\right]^{\tau\tau'}_{l_1l_2,l_3l_4} \rho^{\varsigma}_{\tau'l_4l_3}
\label{eq:S_loc_fin_app
}
\end{align}

\begin{figure}[t!]
\includegraphics[width=0.8\linewidth]{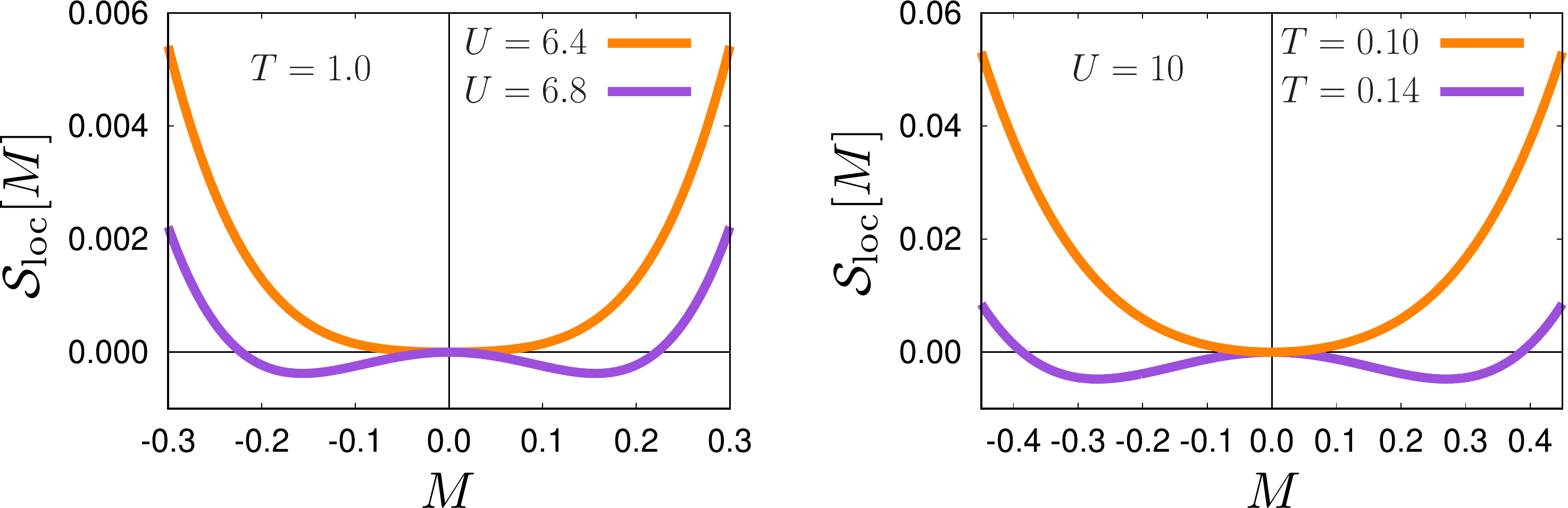}
\caption{\label{fig:energy_app} Local free energy ${\cal S}_{\rm loc}[M]$ of a single band Hubbard model as a function of the local magnetic moment $M$. Results on the left panel are obtained for $T=1$ at different interaction strengths $U=6.4$ (orange line) and $U=6.8$ (violet line). Results on the right panel are obtained for $U=10$ for different temperatures $T=0.10$ (orange line) and $T=0.14$ (violet line).}
\end{figure}

Figure~\ref{fig:energy_app} illustrates the local free energy for a single band Hubbard model considered in the main text as a function of the local magnetic moment ${M_{\omega=0}}$. 
Left panel shows the results obtained for a fixed temperature ${T=1}$ for different values of the Coulomb interaction ${U=6.4}$ (orange line) and ${U=6.8}$ (violet line) around the transition point $U\simeq6.5$. 
Results on the right panel are obtained for a fixed value of the Coulomb interaction $U=10$ for different temperatures $T=0.10$ (orange line) and $T=0.14$ (violet line) below and above the transition point $T\simeq0.12$. 
Crossing the transition point the form of the free energy changes from a paraboloid-like to a double-well potential or {\it vice versa}.\\

\twocolumngrid
\section{Equation of motion for spin degrees of freedom}
\label{app:eq_motion}

In this section we derive the equation of motion for the local magnetic moment.
After leaving out the fast degrees of freedom in Eq.~\eqref{eq:S6} we are left with an action depending on angle variables only. It reads:
\begin{gather}
\mathcal{S}_{\text{spin}} = iS \int^{\beta}_0 d\tau \sum_i\dot{\varphi_i}(1-\cos{\theta}) 
- S\int_{0}^{\beta} d\tau \sum_{i,s} e^{s}_{i\tau} h^{{\rm soc}\,s}_{i\tau} \notag\\
+ 2S^2\iint^{\beta}_0 d\tau \, d\tau' \sum_{ij,ss'}\mathcal{I}^{ss'}_{ij}(\tau-\tau') e^s_{i\tau}e^{s'}_{j\tau'} 
\label{spin_action}
\end{gather}
Here $S$ accounts for fast degrees of freedom associated with local moment value, $\mathcal{I}_{ij}$ is the effective exchange given by Eq.~\eqref{eq:exch}, and $h^{{\rm soc}\,s}_{j\tau}$ is an effective magnetic field introduced in Eq.~\eqref{eq:h_soc}. 
At this point we make use of the fact that the exchange is determined by the super-exchange processes due to electrons and thus has a ``fast'' dependence on time, while the time-dependence of the angle variables is ``slow''. 
This allows one to replace the $\tau'$ time argument of $e^{s'}_j$ with $\tau$ and evaluate the $\tau'$-integration of $\mathcal{I}^{ss'}(\tau-\tau')$ to obtain the zero frequency Fourier component $\mathcal{I}^{ss'}(\omega=0)$. 
We will discuss the corrections to this approximation below.

Now we vary the resulting spin action with respect to $\varphi_i$ and $\theta_i$. As we consider only the first variation of the action variables at different sites can be varied independently, \emph{i.e.} the equation of motion for the spin at a given site is obtained as if all other spins are kept constant. Also note that within our approach $\mathcal{I}^{ss'}_{ij}$ is purely non-local as the introduced Green's function for the fields $f^{(*)}$~\eqref{eq:G_dual_app} is non-local. Thus we can perform the sum over $j$ in the last term of Eq.~\eqref{spin_action} to get:
\begin{align}
\mathcal{S}_{\text{spin}} = \int^{\beta}_0 d\tau \sum_{i} \Bigg\{ iS \dot{\varphi_i}(1-\cos{\theta}) - S \vec{e}_{i\tau}\vec{h}_{i\tau} \Bigg\} 
\label{spin_action_equal_tau}
\end{align}
where ${\vec{h}_{i\tau}=\vec{h}^{0}_{i\tau} + \vec{h}^{\rm soc}_{i\tau}}$, and $\vec{h}^{0}_{i\tau}$ is an effective time-dependent magnetic fields due to interaction with all other spins with the components  
\begin{align}
h^{0s}_{i\tau} &= -~2S\sum_{j,s'} \left[\mathcal{I}^{ss'}_{ij}(\omega=0) + \mathcal{I}^{s's}_{ji}(\omega=0) \right]e^{s'}_{j\tau} \notag\\
&=-~4S\sum_{js'}\mathcal{I}^{ss'}_{ij}(\omega=0) \, e^{s'}_{j\tau}
\end{align}
Calculating the variation of the action with respect to the $i$-th spin is performed straightforwardly by expanding the scalar product in the second term of Eq.~\eqref{spin_action_equal_tau} and yields:
\begin{align}
\delta_i\mathcal{S}_{\text{spin}} &= iS\int^{\beta}_0 \left(\delta\dot{\varphi}_i(1-\cos{\theta_i})+\sin{\theta_i}\delta\theta_i\dot{\varphi_i}\right) \notag\\
&-Sh_{ix}\int^{\beta}_0(\cos{\theta_i}\delta\theta_i\cos{\varphi_i}-\sin{\theta_i}\sin{\varphi_i}\delta\varphi_i) \notag\\
&-Sh_{iy}\int^{\beta}_0(\cos{\theta_i}\delta\theta_i\sin{\varphi_i}+\sin{\theta_i}\cos{\varphi_i}\delta\varphi_i) \notag \\ 
&+Sh_{iz}\int^{\beta}_0\sin{\theta_i}\delta\theta_i
\end{align} 
Now, applying integrating by parts, grouping up terms proportional to $\delta\varphi_i$ and $\delta\theta_i$, and equating the variation to zero we obtain following equations of motion (the index $i$ is omitted as equations for all sites are identical):
\begin{align}
i\sin{\theta}\,\dot{\theta}&=h_x\sin{\theta}\sin{\varphi}-h_y\sin{\theta}\cos{\varphi}\\
i\sin{\theta}\,\dot{\varphi}&=h_x\cos{\theta}\cos{\varphi}+h_y\cos{\theta}\sin{\varphi}-h_z\sin{\theta}
\end{align}
Substituting these two equation into the expression for the time derivatives of $\vec e_{i\tau}$ components
\begin{align}
\dot{\vec{e}}_{i\tau}\equiv\Bigg(&\cos{\theta}\cos{\varphi}\,\dot{\theta}-\sin{\theta}\sin{\varphi}\,\dot{\varphi}; \notag\\
&\cos{\theta}\sin{\varphi}\,\dot{\theta}+\sin{\theta}\cos{\varphi}\,\dot{\varphi};-\sin{\theta}\,\dot{\theta}\Bigg),
\end{align}
and applying simple trigonometric identities we obtain
\begin{align}
i\dot{\vec{e}}_{i\tau}=-\Bigg(&h_y\cos{\theta}-h_z\sin{\theta}\sin{\varphi};\notag\\
&h_z\sin{\theta}\cos{\varphi}-h_x\cos{\theta}; \notag\\
&h_x\sin{\theta}\sin{\varphi}-h_y\sin{\theta}\cos{\varphi}\Bigg) \notag\\
&\hspace{-0.8cm}=-\,\vec{h}_{i\tau}\times\vec{e}_{i\tau}
\label{eq_motion}
\end{align}
By changing $\tau\longrightarrow it$ we restore the Landau-Lifshitz equation.

Now let us discuss the effect of non-equal times in Eq.~\eqref{spin_action}. When varying the exchange term we get two terms, one from varying each of $\vec{e}$. In the second term we can interchange the summation and integration variables $i,s,\tau\longleftrightarrow j,s',\tau'$, and make use of symmetry $\mathcal{I}^{ss'}_{ij}(\tau)=\mathcal{I}^{s's}_{ji}(-\tau)$. This leads to the same result~\eqref{eq_motion} but with a different expression for the effective field:
\begin{equation}
h^{s}_{i\tau} = -~4S\int_0^{\beta} d\tau'\sum_{j,s'}\mathcal{I}^{ss'}_{ij}(\tau-\tau') \, e^{s'}_{j\tau'} + h^{{\rm soc}\,s}_{i\tau}
\end{equation}
So in the general case equation of motion for the magnetic moment is a set of integro-differential equations. 
As the effective exchange is a fast function of time, we can expand the spin time-dependence in powers of $\tau-\tau'$.
The first order term would lead to a term proportional to the time derivative of the spin in the effective field, which would be the Gilbert damping. 
But this term vanishes in the Matsubara time due to the exchange being an even function of time. 
This is not really surprising as one can not expect dissipation effects to be visible in the equilibrium formalism. 
On the other hand, if at this point we perform analytical continuation to real times the Matsubara exchange transforms to a retarded function $I^{{\rm R}\,ss'}_{ij}(t-t')$ and the contribution with the first order time derivative of the spin does not vanish. 
Up to this order we can write~\cite{Sayad_2015}:
\begin{align}
h^{s}_{i}(t) = &-4S\sum_{j,s'} I^{\mathrm{R}\,ss'}_{ij}(\Omega=0) \, e^{s'}_{j}(t) + h^{{\rm soc}\,s}_{i}(t) \notag\\
&-4S\sum_{j,s'}\frac{\partial}{\partial\Omega}\left. {\rm Im} \, I^{\mathrm{R}\,ss'}_{ij}(\Omega)\right|_{\Omega=0}\dot{e}^{s'}_{j}(t)
\end{align}
The last term accounts for the Gilbert damping. $I^{{\rm R}\,ss'}_{ij}(\Omega)$ is a Fourier transform of the retarded exchange interaction $I^{{\rm R}\,ss'}_{ij}(t-t')$ to real frequency $\Omega$.

\section{Approximate criterion for the formation of the local magnetic moment}
\label{app:Sigma}

In this section we relate the criterion for the formation of the local magnetic moment to the self-energy of electrons. 
As has been shown in the Ref.~\cite{PhysRevLett.121.037204}, the fermionic frequency dependence of the four-point vertex and, as a consequence, of the three-point vertex can be neglected provided that collective electronic fluctuations in the corresponding (charge, spin etc.) channel are strong.  
In this case the local exchange interaction~\eqref{eq:J_loc} written in Matsubara frequency space can be approximated as
\begin{align}
J^{\rm loc}_{\omega} &= 
\sum_{\nu,\sigma} \Lambda^{\hspace{-0.05cm}s}_{\nu\omega} \, g^{\sigma}_{\nu} g^{\sigma}_{\nu+\omega} \Lambda^{\hspace{-0.05cm}*\,s}_{\nu\omega} \notag\\
&\simeq \Lambda^{\hspace{-0.05cm}s}_{\av{\nu}\omega} \sum_{\nu,\sigma}  g^{\sigma}_{\nu} g^{\sigma}_{\nu+\omega} \Lambda^{\hspace{-0.05cm}*\,s}_{\nu\omega}
\end{align}
where we introduced a fermionic-frequency-independent three-point vertex $\Lambda^{\hspace{-0.05cm}s}_{\av{\nu}\omega}$.
Using the definition of the three-point vertex function~\eqref{eq:Lambda_app}, one gets the following expression for the approximate local exchange interaction
$
J^{\rm loc}_{\omega}
\simeq \Lambda^{\hspace{-0.05cm}s}_{\av{\nu},\omega},
$
which corresponds to the renormalized local coupling between the magnetic moment and electronic degrees of freedom of the fermion-boson problem~\eqref{eq:S4}. 
The three-point vertex at the zeroth Matsubara frequency ${\omega=0}$ can be obtained from the variation of the self-energy with respect to the local magnetic moment $M_{\omega=0}$ according to Eq.~\eqref{eq:vertex_sigma}. 
Therefore, the criterion for the formation of the local magnetic moment~\eqref{eq:local_condition} can be rewritten as
\begin{align}
{\cal C}_{\omega=0} &= \left(\Pi^{s\,{\rm imp}}_{\omega=0}\right)^{-1} - J^{\rm loc}_{\omega=0} \notag\\
&\simeq \left(\Pi^{s\,{\rm imp}}_{\omega=0}\right)^{-1} - \Lambda^{\hspace{-0.05cm}s}_{\av{\nu},\omega=0} \notag\\
&= \left(\Pi^{s\,{\rm imp}}_{\omega=0}\right)^{-1} - \left(\chi^{s}_{\omega=0}\right)^{-1} - \partial_{M}\Sigma^{s\,{\rm imp}}_{\av{\nu}} \notag\\
&= U^{s} - \partial_{M}\Sigma^{s\,{\rm imp}}_{\av{\nu}} \notag\\
&= - \partial_{M}\overline{\Sigma}^{s\,{\rm imp}}_{\av{\nu}} = 0
\label{eq:local_condition_app}
\end{align}
In this expression ${\overline{\Sigma}^{s\,\rm imp}_{\nu} = \Sigma^{s\,\rm imp}_{\nu} - \Sigma^{\rm H}}$ is the self-energy without the static Hartree contribution ${\Sigma^{\rm H} = -\frac12U\av{M}}$, where ${U^{s} = -U/2}$ is the bare interaction in the spin channel.
The presence of the local magnetic moment in the system is signalled by the positive value of the local self-exchange interaction ${{\cal C}_{\omega=0}>0}$.
The later corresponds to the negative value of the variation of the self-energy ${\partial_{M}\overline{\Sigma}^{s\,{\rm imp}}_{\av{\nu}}<0}$, which describes how electrons at a given lattice site react on the formation of the local magnetic moment. 
In particular, the negative value of the variation indicates that the presence of the local moment in the system is energetically favorable, because it minimizes the self-energy of the electrons. 

\begin{figure}[t!]
\includegraphics[width=1.0\linewidth]{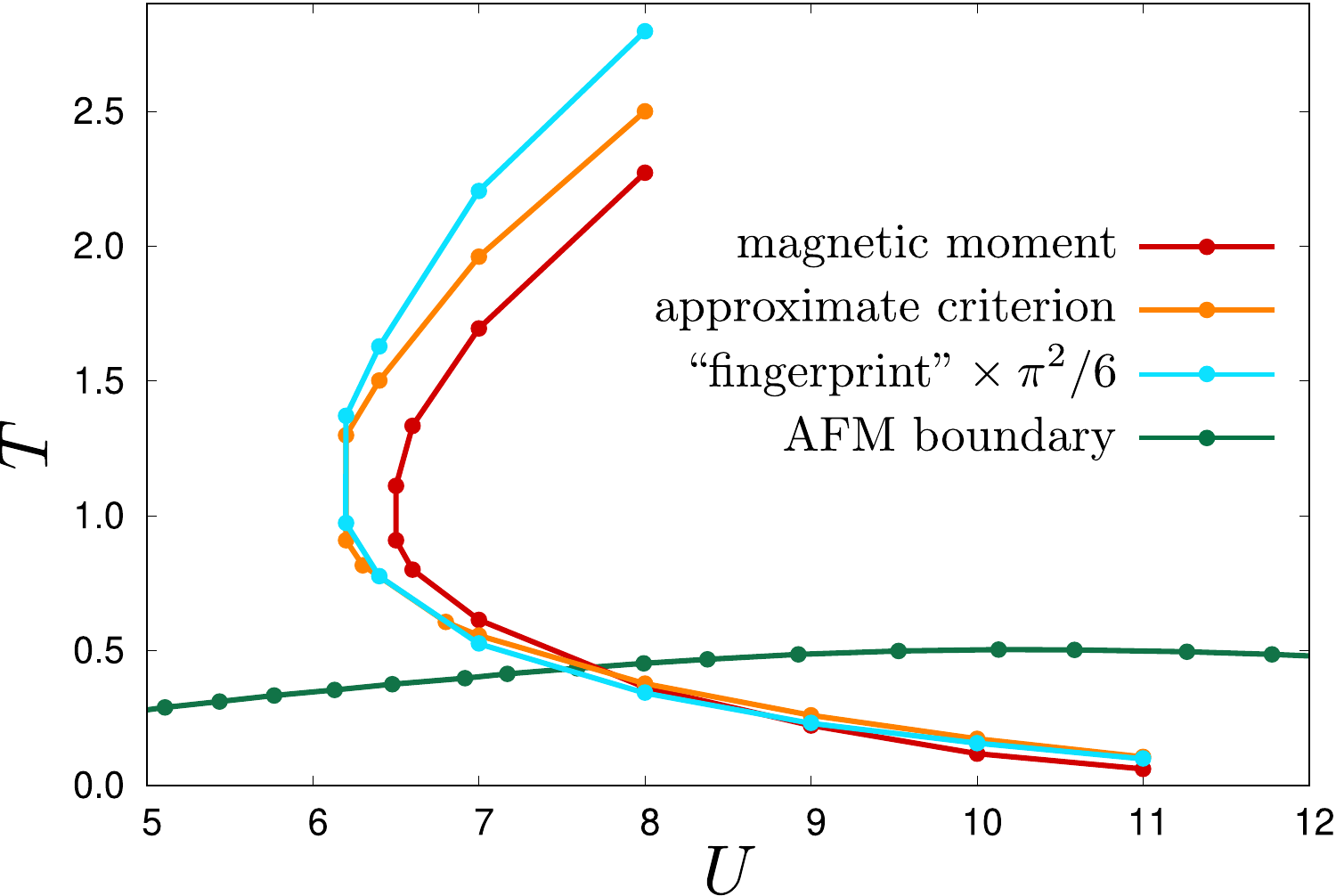}
\caption{\label{fig:phase_app} Phase diagram for the 3D Hubbard model as a function of temperature $T$ and local Coulomb interaction $U$. Red and orange lines correspond to the exact~\eqref{eq:local_condition} and the approximate~\eqref{eq:local_condition2_app} criteria for the formation of the local magnetic moment, respectively. Light blue line depicts the ``fingerprint'' condition proposed in Ref.~\cite{PhysRevLett.126.056403} (see main text) multiplied by a factor of $\pi^2/6$. Green line is the AFM phase transition boundary predicted by DMFT~\cite{PhysRevB.92.144409}.}
\end{figure}

As follows from Eq.~\eqref{eq:local_condition_app}, the approximate condition for the formation of the local magnetic moment can also be written as
\begin{align}
\Lambda^{\hspace{-0.05cm}s}_{\av{\nu},\omega=0} = \left(\Pi^{s\,{\rm imp}}_{\omega=0}\right)^{-1}
\label{eq:local_condition2_app}
\end{align}
We note that the inverse of the polarization operator that stays at the right-hand side of this equation is the high-frequency asymptotics of the three-point vertex function~\cite{PhysRevLett.121.037204, PhysRevB.100.205115}
\begin{align}
{\Lambda^{\hspace{-0.05cm}s}(\nu\to\infty,\omega)= \left(\Pi^{s\,{\rm imp}}_{\omega}\right)^{-1}}
\end{align}
Therefore, it is convenient to use the zeroth fermionic frequency ${\nu_0 = \pi/\beta}$ to approximate the fermionic-frequency-independent three-point vertex
\begin{align}
{\Lambda^{\hspace{-0.05cm}s}(\av{\nu},\omega=0)\simeq\Lambda^{\hspace{-0.05cm}s}(\nu_0,\omega=0)}
\label{eq:vertex_approx_app}
\end{align}
As Fig.~\ref{fig:phase_app} shows, the orange line that corresponds to the approximate condition~\eqref{eq:local_condition2_app} with the approximation~\eqref{eq:vertex_approx_app} for the vertex lies close to the red line that was obtained using the exact criterion for the formation of the local magnetic moment~\eqref{eq:local_condition}.
It is also interesting to note that in the low-temperature regime the result obtained via the ``fingerprint'' condition~\cite{PhysRevLett.126.056403} being multiplied by the scaling factor of $\sum_{n=1}^{\infty} 1/n^2 = \pi^2/6$ (light blue line in Fig.~\ref{fig:phase_app}) is in a very good agreement with the result obtained via the approximate criterion~\eqref{eq:local_condition2_app}.

\bibliography{Ref}

\end{document}